\documentclass[%
superscriptaddress,
onecolumn,
notitlepage,
aps,
prb,
11pt
]{revtex4-2}

\usepackage[font=small]{caption}
\usepackage[labelformat=empty, position=top]{subcaption}
\usepackage{graphicx}
\usepackage[export]{adjustbox}
\usepackage[utf8]{inputenc}
\usepackage{array}
\usepackage{diagbox}
\usepackage{booktabs}
\usepackage{mathrsfs}
\usepackage{makecell}
\usepackage{siunitx}
\usepackage{tikz}
\usepackage{amsmath}
\usepackage{amsthm}
\usepackage{amssymb}
\usepackage{multirow}
\usepackage[rightcaption]{sidecap}
\usepackage{wrapfig}

\usepackage[hidelinks]{hyperref}
\setcitestyle{super}

\begin{document}

\title{Global properties of the energy landscape: a testing and training arena for machine learned potentials}

\author{Vlad C\u{a}rare}
\email{vcarare@xyme.ai}
\affiliation{Xyme, UK}
\affiliation{Yusuf Hamied Department of Chemistry, University of Cambridge, UK}
\affiliation{IBM Research Europe, Daresbury, UK}

\author{Fabian L.~Thiemann}
\email{fabian.thiemann@ibm.com}
\affiliation{IBM Research Europe, Daresbury, UK}

\author{Joe D.~Morrow}
\email{joe@cusp.ai}
\affiliation{IBM Research Europe, Daresbury, UK}

\author{David J.~Wales}
\email{dw34@cam.ac.uk}
\affiliation{Xyme, UK}
\affiliation{Yusuf Hamied Department of Chemistry, University of Cambridge, UK}

\author{Edward O.~Pyzer-Knapp}
\email{ed@xyme.ai}
\affiliation{Xyme, UK}

\author{Luke Dicks}
\email{Corresponding author: ldicks@xyme.ai}
\affiliation{Xyme, UK}
\affiliation{IBM Research Europe, Daresbury, UK}

\date{\today}

\begin{abstract}

Machine learning interatomic potentials (MLIPs) have achieved remarkable accuracy on
standard benchmarks, yet their ability to reproduce molecular kinetics --
critical for reaction rate calculations -- remains largely unexplored. We
introduce Landscape17, a dataset of complete kinetic transition networks (KTNs)
for the molecules of the MD17 dataset, computed using hybrid-level density functional
theory. Each KTN contains minima, transition states, and approximate steepest-descent
paths, along with energies, forces, and Hessian eigenspectra at stationary
points. We develop a comprehensive test suite to evaluate the MLIP ability to
reproduce these reference landscapes and apply it to a number of state-of-the-art architectures.
Our results reveal limitations in current MLIPs: all the models considered miss over
half of the DFT transition states and generate stable unphysical structures
throughout the potential energy surface. 
Data augmentation with pathway configurations improves
reproduction of DFT potential energy surfaces, resulting in significant
improvement in the global kinetics. However, these models still produce many spurious
stable structures, indicating that current MLIP architectures face underlying
challenges in capturing the topology of molecular potential energy surfaces.
The Landscape17 benchmark provides a straightforward but demanding test of MLIPs
for kinetic applications, requiring only up to a few hours of compute
time. We propose this test
for validation of next-generation MLIPs targeting reaction discovery and rate prediction.

\end{abstract}


\maketitle

\section{Introduction}

Physics-based approaches to molecular simulation have long
served as the foundation for \textit{in silico} molecular investigation and
discovery.  Expanded computational power has enabled 
these tools to answer sets of increasingly complex questions, with
notable success.\cite{pyzer-knapp_advancing_2024}

Over the past decades, machine learning interatomic potentials (MLIPs) have emerged as an
alternative component for molecular simulations. These models -- often neural
networks – are trained on high-quality reference data and hold the promise of
providing \textit{ab initio} accuracy at a fraction of the computational
cost.\cite{deringer_machine_2019,Unke2021, behler_four_2021, thiemann_introduction_2025} This efficiency provides access to longer simulation timescales, and
the computation of a wide range of derived properties.

Various architectures\cite{schutt_schnet_2018,
ANI,qiao_orbnet_2020,unke_spookynet_2021,gasteiger_directional_2022,
Nequip, MACE, Allegro, ani2x, Aimnet2} and molecular representations\cite{bartok_gaussian_2010, Behler2011,Drautz2019,chmiela_sgdml_2019,Huo2022} have been developed for the
construction of MLIPs. These models are trained on energies, forces and, for
periodic materials, stresses, which are most commonly computed using density
functional theory (DFT). Foundational
models have been proposed for
materials\cite{mace-mp-0b3,rhodes_orb-v3_2025,fu_learning_2025}
and molecules\cite{kovacs_mace-off_2025,
kabylda_molecular_2025,frank2024euclidean,wood_uma_2025} based on large-scale
datasets,\cite{deng_chgnet_2023, eastman_spice_2023,levine_open_2025,
ramakrishnan_quantum_2014} which aim to improve out-of-distribution
generalizability, potentially offering improved representations for system-specific
applications.

Underlying such advances are both comprehensive
benchmarks\cite{matbench_riebesell_framework_2025} and lightweight tests that
allow for on-the-fly tracking of architectural or training dataset changes. One
example is the popular rMD17 benchmark,\cite{MD17, rMD17} which provides energy
and force labels for samples obtained from \textit{ab initio} molecular
dynamics (MD) trajectories of small molecules. Given its simplicity and the
desire for models to faithfully sample from the Boltzmann distribution, rMD17
has been used as a first benchmark for model accuracy.\cite{MACE,
Nequip, Allegro,pelaez_torchmd-net_2024}

While valuable, standard MD simulations do not address broken ergodicity problems.
At low temperature they tend
to sample low-diversity, low-energy states around local energy minima, 
trapped between energy barriers that restrict exploration of
the complete potential energy surface (PES) on the accessible MD time scale. Increasing temperature can help
overcome such barriers, but may then
miss the configurations and pathways essential for accurate molecular kinetics
calculations at the temperatures of interest (e.g.~for reaction rates).\cite{curse_of_dimensionality,Staub2023,Shenoy2023,Swinburne2020,Kong2025}
These bottlenecks can also compromise thermodynamic predictions when competing low-energy
minima remain unobserved due to insufficient sampling.
Consequently, due to its relatively low diversity,
models released over the past years have saturated the energy and force errors
on the rMD17 benchmark.\cite{MACE, Nequip, Allegro,pelaez_torchmd-net_2024}

While enhanced sampling methods, such as parallel tempering\cite{HukushimaN96,TesiROW96,Hansmann97}
MD replica exchange,\cite{SugitaO99}
multicanonical,\cite{NakajimaNK97}
Wang-Landau,\cite{WangL01}
and biasing schemes,\cite{torriev74,LaioP02} can help to overcome broken ergodicity problems, the computational expense limits routine
application for generating large databases. 
Methods to explicitly treat rare event dynamics\cite{BolhuisCDG02,VanErpMB03,FaradjianE04,AllenFW06b,ChoderaDSPSP07,DellagoB09,PandeBB10,PrinzWSKSHCSN11,SwinburneP18} are also computationally demanding.
An alternative approach is to construct a kinetic transition network\cite{NoeF08,pradag09,Wales10a}
using discrete path sampling,\cite{Wales02,Wales04}
employing methods based on geometry optimization to characterize local minima and the pathways that
connect them via transition states in the potential energy surface.
Here, a transition state is defined as a stationary point with a single negative Hessian eigenvalue.\cite{MurrellL68}
Global kinetic properties are then extracted using unimolecular rate theory for the individual
minimum-to-minimum rates.\cite{forst73,Laidler87}
This formalism involves solution of a master equation\cite{kampen81} for the 
global dynamics, as for dynamics-based schemes that produce a Markov state representation.\cite{ChoderaDSPSP07,BowmanBBP09,PrinzWSKSHCSN11,LaneBBVPTY11}

To address the need for kinetically-relevant structures in developing MLIPs, 
specialized datasets have been developed that capture transition paths. 
Examples include Transition1x,\cite{transition1x} which features samples from
nudged-elastic-band (NEB) constructions across various reactions, and more
comprehensive, data-rich collections such as OMol25\cite{levine_open_2025} and
SPICE.\cite{eastman_spice_2023}
The latter examples, created as training sets for MLIPs, incorporate more
extensive sampling of molecular conformations, including specialized subsets
that capture configurations along transition state pathways. 
Automated transition state finding is an important objective of many studies,
and recent works explore the applicability of generative models:
SDiff,\cite{tsdiff} TSGen,\cite{tsgen} and React-OT,\cite{reactot} as well as
workflows to improve MLIPs performance in NEBs.\cite{kuryla_efficient_2025}

Given this context, there is a need for lightweight benchmarks that assess
both the accuracy and transferability of MLIPs across the kinetically-relevant
paths in the energy landscape. Such a benchmark should allow the
validation of MLIPs beyond energy and force errors to include the
organization of the potential energy landscape. In particular, we wish
to reproduce stationary points faithfully, without
additional spurious local minima (or transition states). We present two major
contributions in this direction.

Firstly, we introduce the Landscape17 dataset, which systematically expands
upon rMD17 by providing complete kinetic transition networks
(KTNs)\cite{NoeF08,pradag09,Wales10a} for the six molecules within rMD17 that have more than one
distinct local minimum structure. This dataset features global potential energy
surface representations generated using the energy landscape
framework,\cite{Wales2003,Wales18} and includes regions crucial for accurately
reproducing both thermodynamic and kinetic properties. For each of the selected
six molecules (ethanol, malonaldehyde, paracetamol, salicylic acid, azobenzene, and
aspirin) we provide all the minima and transition states, along with configurations from the two approximate
steepest-descent paths connecting each transition state to the corresponding
minima. These paths underpin the most probable routes between minima at finite temperature, offering
essential configurations for understanding system kinetics.

Secondly, we establish a comprehensive, yet lightweight, testing suite for
evaluating energy landscape fidelity, and apply this to assess
current MLIP architectures. The ability of MLIPs to capture entire reference DFT
KTNs, with the corresponding energies and atomic structures of both minima and
transition states, extends molecular validation metrics beyond conventional
force and energy errors. Since PES organization strongly determines physical
properties, reproducing these properties provides an appropriate
and generalized test of overall physical property reproduction.

Our analysis reveals that current ML models struggle to accurately predict KTNs
for these small molecules. Moreover, we find that current MLIPs often exhibit unphysical minima,
an issue that is not usually a problem for semi-empirical
methods such as GFN2-xTB.\cite{bannwarth_gfn2-xtbaccurate_2019} We demonstrate
systematic improvement strategies using configurations sampled along the
pathways, and suggest that this dataset and benchmark suite constitute powerful tools
for evaluating model potentials.

\section{Results}

In this contribution, we extend the commonly used rMD17 benchmark dataset to produce
Landscape17, which provides KTNs for the six molecules within rMD17 that
exhibit multiple distinct minima: ethanol, malonaldehyde, salicylic acid,
azobenzene, paracetamol, and aspirin. These networks were computed using
hybrid-functional DFT, at an estimated computational cost exceeding 10$^5$ CPU
hours, and capture the pathways that are essential for a proper description of global kinetics.
The dataset includes configurations from the pathways to supplement the stationary points and is publicly available.\cite{landscape17dataset}

We benchmark several MLIP architectures for the selected 
molecules to evaluate how well they reproduce key features of the reference DFT
potential energy surfaces, and investigate whether incorporating
pathway data improves MLIP performance. We begin with a
brief overview of the DFT potential energy surfaces and data acquisition
regimes, followed by an evaluation of MLIPs trained using the corresponding
datasets. We then examine the resulting MLIP surfaces, and compare with
the DFT reference.

\subsection{DFT landscapes}

To generate DFT KTNs, we follow an established procedure that begins with
basin-hopping global optimization\cite{lis87,Wales1997,waless99} to identify low-energy minima, followed by
combined single- and double-ended searches to locate transition states;
these methods have been explained in detail elsewhere.\cite{Wales2003,JosephRCMW17,Wales18,RoederJHW19}
Here, transition states are defined geometrically as points of (approximate) zero gradient and
exactly one negative Hessian eigenvalue.\cite{MurrellL68} We utilized TopSearch,\cite{topsearch}
an open-source Python package developed by some of the authors, to perform
landscape exploration. For each KTN, we excluded repeated permutational isomers
and structures related by the inversion operation, as these can be
reconstructed through symmetry operations. Further details of the landscape
generation process are available in the Methods section and Appendix
\ref{sec:appendix-landscape17}.

Table \ref{tab:landscape_data} presents the number of minima and transition
states for the six rMD17 molecules in our study. We also collated data along
approximate steepest-descent paths from transition states to their connected minima,
following small displacements parallel and antiparallel to
the eigenvector associated with the unique negative Hessian eigenvalue.
The complete dataset -- Landscape17 -- includes atomic coordinates, energies,
forces, and Hessian eigenspectra for all the stationary points, along with positions,
energies, and forces for configurations along the pathways. 

\begin{table}[bp]
        \centering
	\caption{Statistics of the Landscape17 dataset, reflecting the counts
of minima and transitions states in the DFT landscape of each molecule. The
number of stationary points includes only distinct structures, lumping together
permutation-inversion isomers.}
        \vspace{0.1cm}
        \begin{tabular}{@{}l*{9}{c}@{}} 
        \toprule
        Molecule & \begin{tabular}[c]{@{}c@{}}\# Min.\end{tabular} & \begin{tabular}[c]{@{}c@{}}\# TS\end{tabular}  \\
        \midrule
        Ethanol & 2 &2  \\
        Malonaldehyde & 2 &4 \\
        Salicylic acid& 7 & 11 \\
        Azobenzene & 2 &4 \\
        Paracetamol & 4 &9  \\
        Aspirin & 11 & 37  \\
        \midrule
        Total & 28 & 67 \\
        \bottomrule
        \end{tabular}
        \label{tab:landscape_data}
        \vspace{-0.3cm}
\end{table}

\begin{figure}[htp!]
        \centerline{\includegraphics[width=1.0\linewidth]{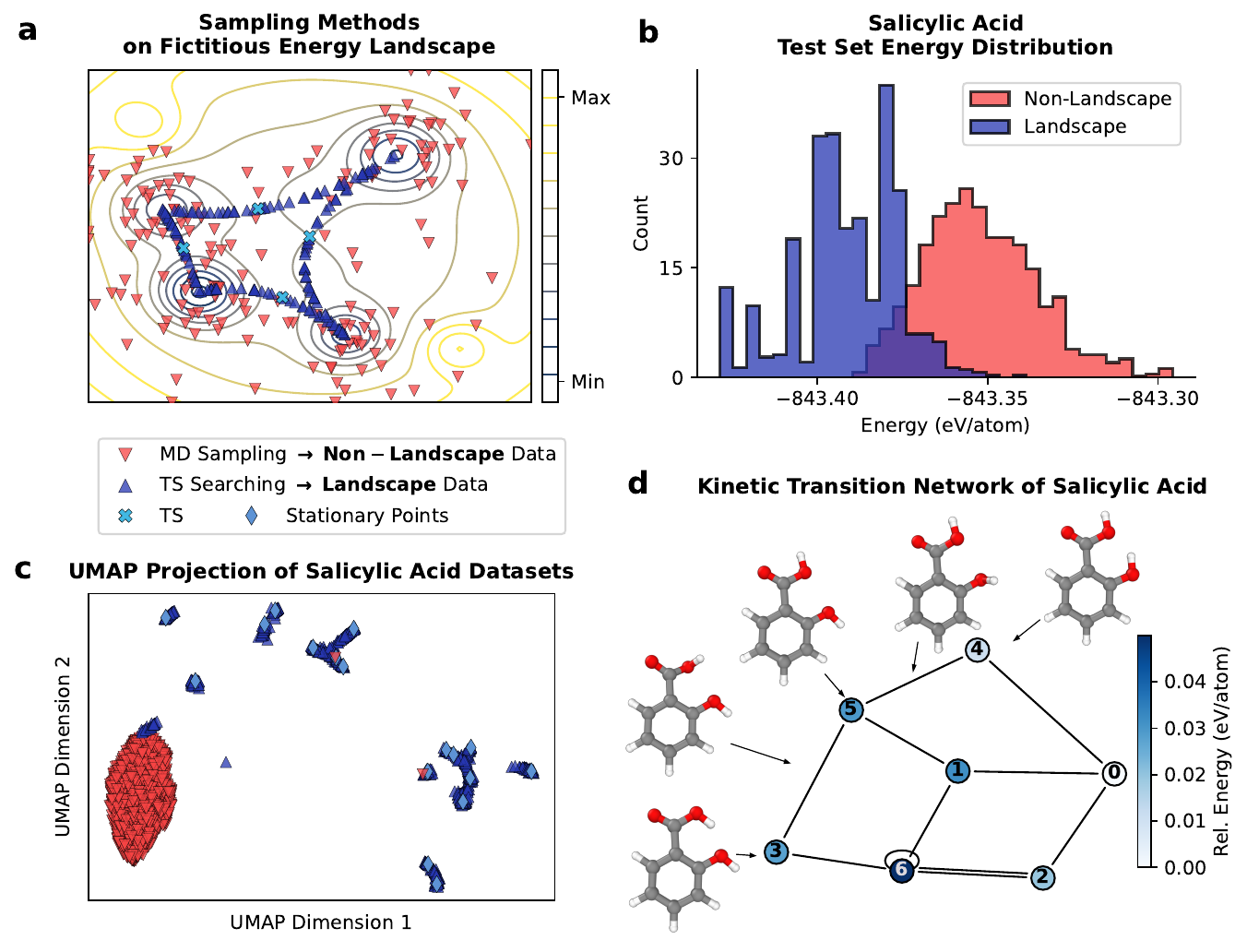}}
	\caption{Contrasting \textbf{Landscape} with \textbf{Non-Landscape}
data. Non-Landscape data is sampled from MD trajectories, while Landscape data
is sampled from approximate paths between transition states and the
connected minima. \textbf{a}) Sketch of the distributions resulting from the
two sampling schemes. \textbf{b}) Energy histogram comparison for test sets of
Non-Landscape and Landscape structures of salicylic acid. The former data was
collected from an MD run at 500\,K (rMD17\cite{rMD17}). \textbf{c}) UMAP\cite{mcinnes_umap_2018} projections of
the structurally-averaged MACE\cite{MACE} descriptors of three salicylic acid datasets,
showing clear distinction between Landscape and Non-Landscape datasets. 
\textbf{d}) Visual representation of the DFT KTN of salicylic acid. Minima are
given by nodes and transition states by edges connecting the nodes. Self-loops
and multi-edges are permitted, representing transition states between degenerate rearrangements
and alternative transition states, respectively.
The colors of the nodes correspond to the energy of the corresponding minima
relative to the global minimum. We highlight the structures along the 3-5-4 pathway.
Images were rendered using OVITO.\cite{ovito}
        \label{fig:fig1}
}
\end{figure}

In this work, we distinguish between two data categories: \textbf{Landscape}
(L) data  from approximate steepest-descent pathways, and \textbf{Non-Landscape} (N-L)
data from molecular dynamics simulations (e.g., rMD17). Landscape data
captures minimum energy paths between stationary points, representing the
essential topology of the potential energy surface, while Non-Landscape data
samples beyond this minimal representation of the connectivity. This distinction is
illustrated in Fig.~\ref{fig:fig1}.
The key difference is that the geometry optimization procedures employed in discrete
path sampling\cite{Wales02,Wales04} can treat both high and low barriers, and the associated long and short time scales,
directly addressing rare event dynamics and broken ergodicity.

Panel \textbf{a} provides a schematic of a hypothetical energy landscape that
contrasts the two sampling methods. Panel \textbf{b} shows energy distributions
for one of the MD17 molecules: salicylic acid. The distributions reveal that
500\,K MD (N-L data) explores higher energy regions than transition state
searching (L data) but produces less diverse distributions, as shown
in panel \textbf{c}. These UMAP\cite{mcinnes_umap_2018} projections of
structurally-averaged MACE\cite{MACE} descriptors further demonstrate a clear separation
between regions sampled by the two methods. Landscape data explores diverse
environments around stationary points, while, in contrast, Non-Landscape data
concentrates in a single region, despite the higher energies.
Panel \textbf{d} presents the DFT
KTN for salicylic acid. KTNs can be visualized as graphs, where nodes represent
minima and edges represent transition states. Self-edges indicate transition
states between minima related by permutation-inversion symmetry;
these are termed degenerate rearrangements.\cite{Nourse80} Similar figures for the
other molecules are provided in Appendix \ref{sec:appendix-datadistr}.

The targeted, discrete path sampling\cite{Wales02,Wales04} approach employed in constructing the Landscape17
dataset, with explicit exploration of transition states, enables
both rigorous benchmarking of these critical areas of the potential energy
surface and systematic analysis of training data effects on model performance.
This sampling strategy provides clear insight
into how different machine learning potential architectures handle the
challenging task of reproducing saddle point regions, which are 
harder to represent accurately due to the inherently unstable eigendirection
with  negative curvature.

\subsection{Performance Impact of Landscape Data Inclusion}

With the Landscape17 dataset we can systematically test whether the inclusion
of landscape-specific data into training sets based on
non-landscape-specific data (such as molecular dynamics trajectories) can help
achieve accurate molecular KTNs. We trained individual models, for
each of the six selected rMD17 molecules, following the protocols established in
the original publications, using the same data across models. The system-specific architectures employed include NequIP,\cite{Nequip} 
MACE\cite{MACE} and Allegro,\cite{Allegro} supplemented
by pretrained foundational models: ANI2x,\cite{ani2x} AIMNet2,\cite{Aimnet2}
MACE-MP-0b3\cite{mace-mp-0b3} and SO3LR;\cite{kabylda_molecular_2025} and the
semiempirical method GFN2-xTB.\cite{bannwarth_gfn2-xtbaccurate_2019} These
models constitute our baseline \textbf{Non-Landscape} (N-L) set. 
Subsequently, we augmented the Non-Landscape training datasets with portions of
the newly generated Landscape17 data (see Methods) to create corresponding
\textbf{Landscape} (L) Allegro, MACE and NequIP models.

\begin{figure}[tp!]
        \centering
        \includegraphics[width=1\linewidth]{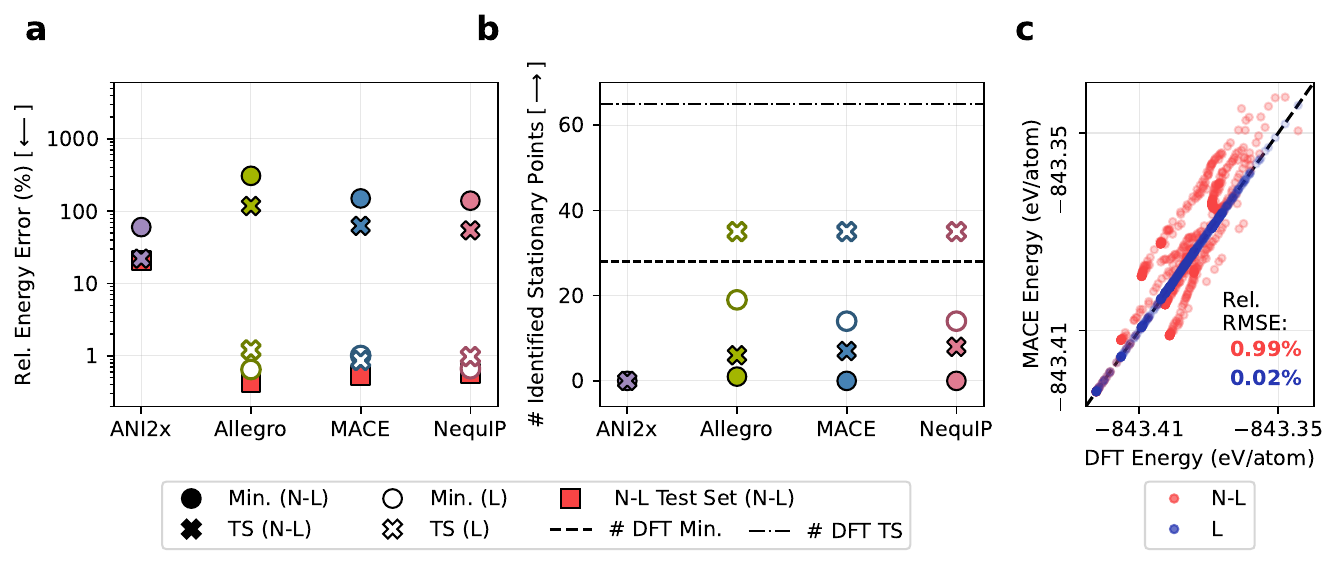}
	\caption{Model performance on the stationary points of the DFT landscapes.
\textbf{a}) Relative energy errors of MLIPs on minima, transition states, and a
MD17 test set, for models trained on Non-Landscape data (N-L models) and the
combination of Non-Landscape \& Landscape data (L models). Values for the N-L
test set and for ANI2x are shown for reference. Relative energy errors are
obtained by dividing the root mean squared errors by the standard deviations of
the ground truth energy distributions. The results are averaged over all six
molecules. \textbf{b}) Number of DFT minima and transition states where the maximum
force components are correctly predicted to be smaller than $10^{-3}$\,eV/Å and
$10^{-2}$\,eV/Å respectively, by the corresponding model. \textbf{c}) Energy
errors parity plot for predictions of Non-Landscape and Landscape MACE models
on a Landscape test set. Error tables for all the other molecules and models
are presented in Appendix \ref{app:training}.}
        \label{fig:FRMSEandERMSE}
\end{figure}

The Landscape models demonstrate superior performance in predicting energies
and forces at critical points on DFT energy surfaces.
Fig.~\ref{fig:FRMSEandERMSE} reveals a consistent improvement across all
architectures, with systematic reductions in both energy and force errors
following landscape data inclusion (with results for ANI2x displayed for
reference). Panel \textbf{a} shows the average relative root-mean-squared
errors (RMSE) for energy predictions on minima and transition states, calculated by
normalizing RMSE values against the standard deviations of target
distributions. Panel \textbf{b} quantifies instances where MLIPs correctly
predict maximum force components below convergence thresholds of
$10^{-3}$\,eV/Å for minima and $10^{-2}$\,eV/Å for transition states - criteria
matching those used for DFT stationary point optimization. Panel \textbf{c}
presents an example energy parity plot comparing MACE N-L and L model
predictions on the aspirin landscape test set. Consistent trends across all
models and molecules are documented through comprehensive MAE and RMSE
statistics provided in the Appendix, in Tables
\ref{table:MAE_and_RMSE_for_non_altitude_training}--\ref{table:MAE_and_RMSE_for_altitude_training_sd_testing}.
It is apparent that N-L models fail, by two orders of magnitude, to accurately predict the energies and forces of stationary points, despite achieving low errors for MD samples, while L models attain low errors on both test sets. This underlines the stark difference between Non-Landscape and Landscape data, even for the small molecules considered here.

\newpage

\begin{figure}[tp!]
        \centering
        \includegraphics[width=1\linewidth]{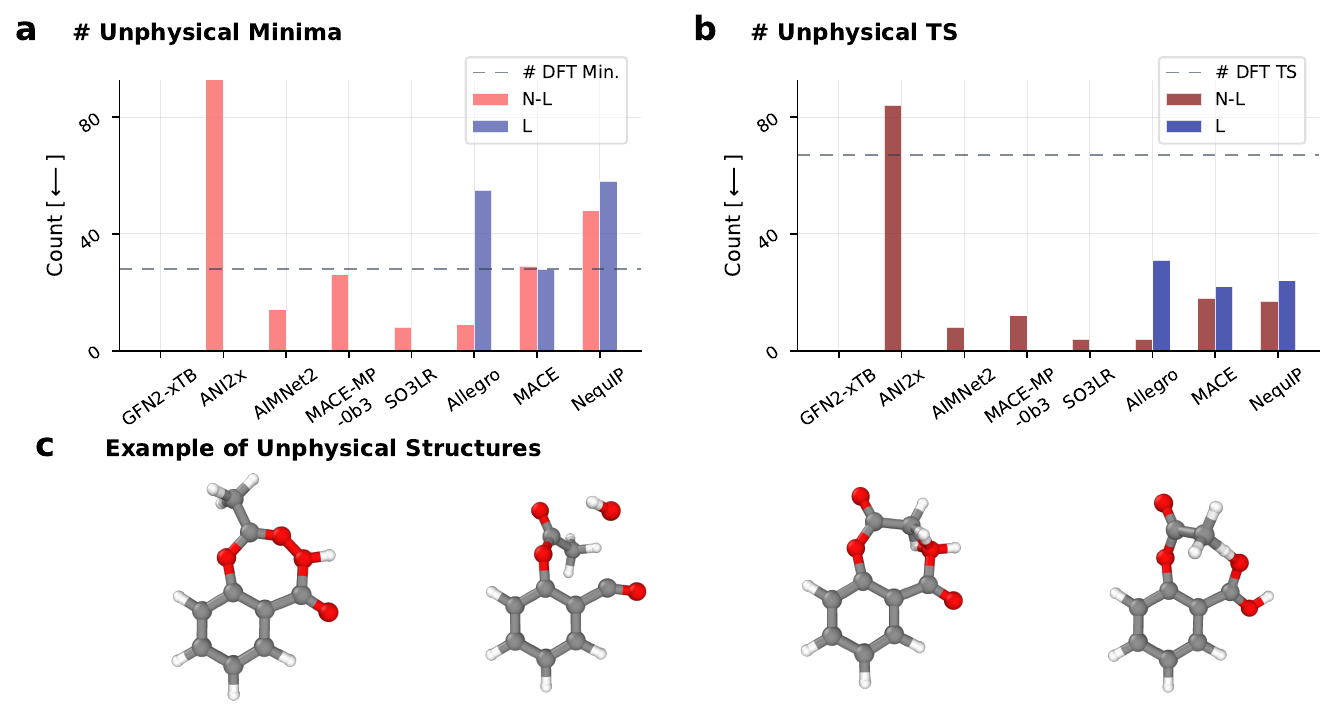}
	\caption{Physicality of MLIP energy landscapes. \textbf{a}) Cumulative counts of
physical and unphysical minima and \textbf{b}) transition states resulting from energy landscape exploration using the
corresponding N-L or L models. \textit{Non-physicality} of a structure is
defined by a difference in its SMILES representation from that of the global
minimum conformation of the respective molecule, or by the presence of a
multiply-bonded hydrogen atom. \textbf{c}) Examples of unphysical
configurations of the aspirin molecule. Images were rendered using
OVITO.\cite{ovito}}
        \label{fig:mlpLandscapes}
\end{figure}

\subsection{Unphysical structures in MLIP landscapes}

To evaluate the performance of machine learning potentials beyond traditional
energy and force metrics, we constructed energy landscapes for each MLIP using
the same algorithms employed for DFT landscape generation. We report KTNs for
each MLIP and molecule after combining results from 20 independent landscape
exploration runs, each initiated from different conformations. Permutation-inversion isomers
are lumped together (see Figs.~\ref{fig:flowchartLandscapeGen} and
\ref{fig:evolutionof_n_ts_n_min} in the Appendix). This multi-run workflow was
not feasible for DFT calculations due to the prohibitive computational cost;
instead, we employed single extended runs followed by thorough manual
enumeration of candidate stationary points (see Methods).

The KTNs for each model were initially assessed through self-consistent
analysis without reference to DFT benchmarks, focusing on the physicality of
discovered stationary points. We define a structure to be unphysical if its
SMILES representation\cite{weininger_smiles_1988} differs from that of the global
minimum, or if it contains hydrogen atoms with multiple bonds.

Our analysis highlights the prevalence of unphysical minima for all MLIPs,
Fig.~\ref{fig:mlpLandscapes}. This phenomenon affects both system-specific
models trained on N-L/L data (Allegro, MACE, NequIP) and foundational models
pretrained on diverse datasets (ANI2x, AIMNet2, MACE-MP-0b3, SO3LR), indicating
a systematic limitation across different MLIP architectures and training
paradigms. It is interesting to note that the effects are magnified for the system-specific models trained on Landscape data, suggesting that adding more training data results in models with rougher energy landscapes. Landscape roughness, and its relation to training data, has been explored in detail elsewhere.\cite{desouza_stevenson_niblett_farrell_wales_2017,Dicks2024-at, Wilson2025} Extended plots for each molecule are shown in
Fig.~\ref{fig:phys-and-non-phys-allmolecs}.

Importantly, our landscape exploration methodology exclusively employs dihedral angle
rotations (see Methods), and application to GFN2-xTB reveals no unphysical
stationary points. Therefore, the prevalence of such structures
serves as a quantitative metric to assess the global MLIP energy landscape. The
emergence of unphysical local minima will hinder
enhanced sampling methods, since these artificial configurations may have
significant statistical weights.

\subsection{Reproduction of reference DFT landscapes by MLIPs}

Faithfully reproducing the KTNs for the underlying quantum chemical level of
theory is a demanding test for machine learned potentials. To evaluate this capability, we removed the unphysical stationary
points from each MLIP landscape and mapped the remaining structures onto
the reference DFT landscape using the algorithms detailed in the
Methods section, and in Appendix,
Figs.~\ref{fig:flowchartOverview}--\ref{fig:flowchartcomparisonAlgorithm}.
These mapping procedures rely on a structure comparison module that calculates
the minimal root-mean-squared-deviation (RMSD) between configurations to
establish correspondence between DFT and MLIP stationary points.

\begin{figure}[tp!]
        \centering
        \includegraphics[width=1\linewidth]{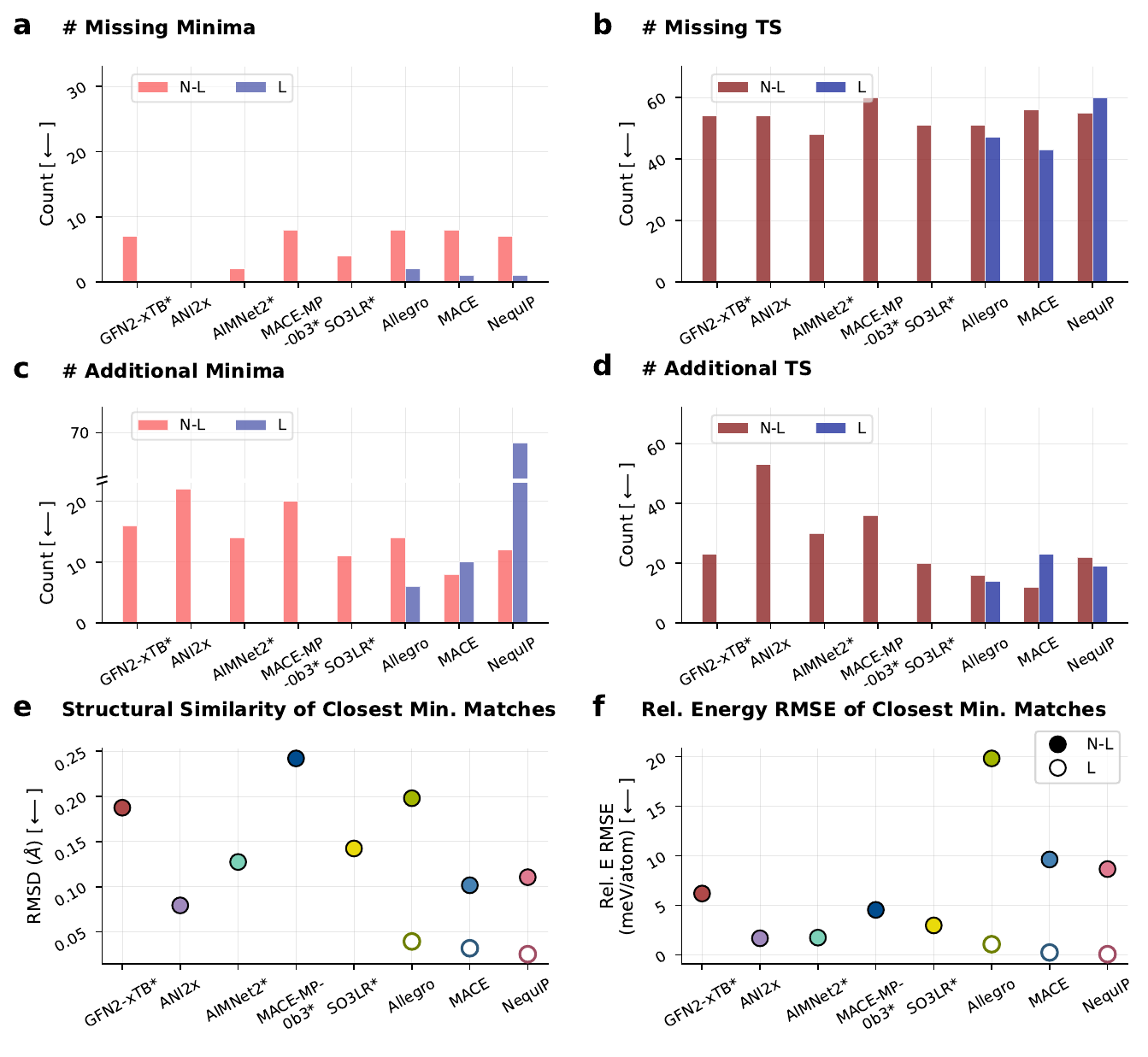}
	\caption{DFT-MLIP KTNs comparison. \textbf{a})--\textbf{d}) \textit{exact} and \textbf{e}), \textbf{f}) \textit{closest} DFT matching statistics for each MLIP KTN. The exact match counts the
number of MLIP stationary points that can (Matched) and cannot (Additional) be
matched to within an RMSD of 0.3\,Å to DFT analogues. The number of Missing minima is computed by subtracting the count of matches from the number of DFT stationary points. The closest match computes the
RMSD and relative energy RMSE of closest MLIP-KTN minima pairs. TS pairs are not
considered because of the inconsistent number of pairings between models. DFT
supports a total of 28 minima and 67 transition states. The asterisk indicates
that energy and forces of the DFT KTN were reevaluated with the DFT functional
matching the training set of the corresponding MLIP.}
        \label{fig:comparison}
\end{figure}

\subsubsection{Reproducing stationary points}

Our comparative analysis addresses two complementary issues: the extent to
which DFT KTN stationary points can be \textit{exactly matched} to MLIP
analogues, and the structural and relative energy accuracy of the
\textit{closest} DFT-MLIP pairs. We define exact matching by an RMSD
threshold of 0.3\,Å for structural equivalence between DFT and MLIP
configurations. For transition state exact matching, we imposed the additional
requirement that the connected minima in the MLIP network must themselves be
exactly matched to corresponding DFT minima. To assess structural and energy
fidelity for the closest MLIP analogues of each DFT minimum, we computed average
RMSD and relative (to the global minimum) energy RMSE values, intentionally
excluding transition states from this analysis due to their systematic
underrepresentation in MLIP landscapes compared to DFT references (see Appendix
Fig.~\ref{fig:phys-and-non-phys-allmolecs}). The results for exact matching are
shown in Fig.~\ref{fig:comparison} \textbf{a}--\textbf{d} (and
Fig.~\ref{fig:all_exact_matches_comparisons}), and for structural and energy fidelity in
\textbf{e} and \textbf{f}.

The result of exact and closest matching demonstrates that incorporating
landscape-specific data into training datasets generally enhances the overlap
between DFT and MLIP KTNs, improving both structures (higher
number of minima matches - panel \textbf{a}, lower RMSD - panel \textbf{e}) and
energetics (lower relative energy RMSE - panel \textbf{f}).
Nevertheless, the challenge of fully reproducing KTNs remains unsolved,
especially for transition states (panel \textbf{b}). Additionally, panels
\textbf{c, d} show that all of the models overpredict the number of stationary points. 
The presence of additional spurious stationary points is a common problem for empirical potentials.\cite{FurmanW19}
For the purpose of studying landscapes where the DFT KTN is not known
\textit{a priori}, it is essential that one chooses a model with the best
trade-off between maximizing exact matches and minimizing artifacts.

The substantial number of successful minima matches and low RMSD values
observed for models trained on data of different DFT functionals (AIMNet2,
MACE-MP-0b3, SO3LR) confirms that the stationary points
remain consistent across functional choices for these six molecules.
This observation helps to validate our suggested benchmarking approach.

\begin{figure}[tp!]
        \centering
        \includegraphics[width=1\linewidth]{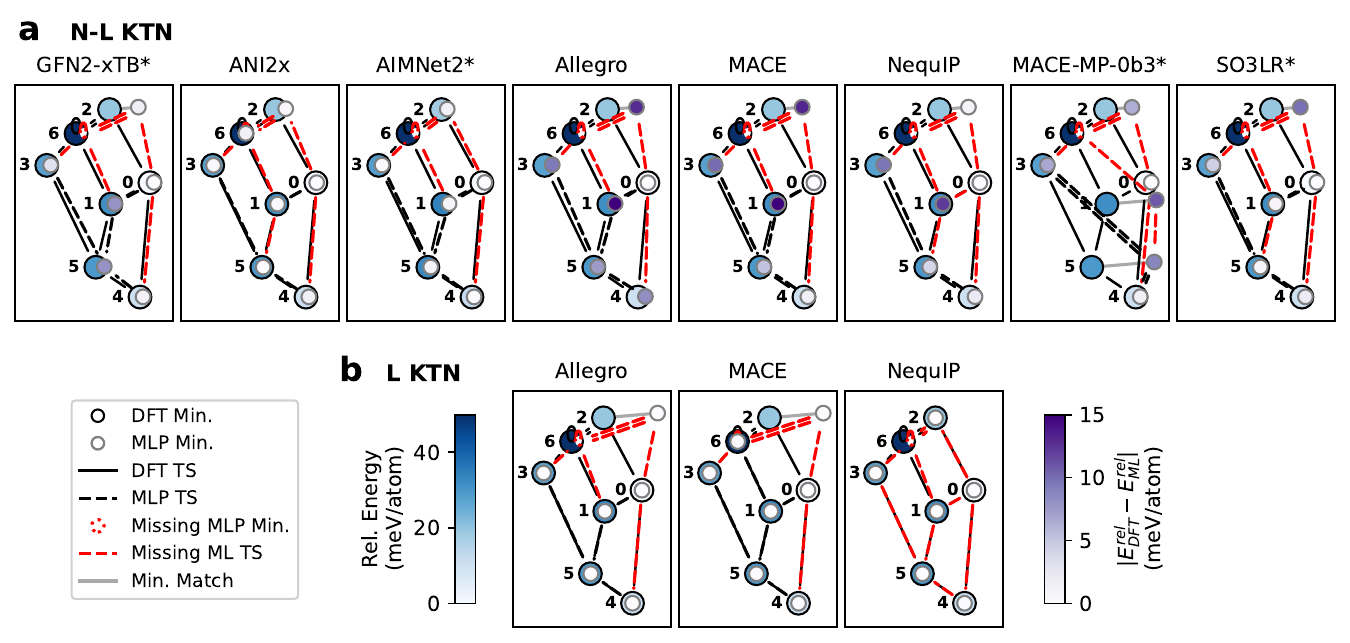}
	\caption{KTNs of MLIPs superimposed on the reference DFT for
salicylic acid. Large circles represent DFT minima and solid black lines
indicate transition states. Distances between DFT minima are scaled
quadratically with the energy barrier between them. Small overlaid circles
represent MLIP minima, with gray contours indicating a matched pair of DFT-MLIP
minima (according to an RMSD threshold of 0.3\,Å), and red indicating a missing
MLIP structure. Matched pairs are linked by gray lines, of distance proportional
to the RMSD between minima at optimal alignment. Black dashed lines represent
transition states in the MLIP KTN which can be mapped to transition states in
the DFT KTN. The correspondence requires successful MLIP-DFT mapping for the
connected minima, as well as structure similarity of 0.3\,Å between the TS. A
red dashed line represent a lack of such a TS. For DFT minima, the colors
indicate the energy difference of the node with respect to the global minimum
(node 0). For MLIP minima, the colors track the energy difference to their
corresponding matched DFT minimum, with white implying a low error.}
        \label{fig:ktncomparison}
\end{figure}

The case of salicylic acid illustrates general trends clearly
(Fig.~\ref{fig:ktncomparison}) when we overlay the MLIP networks on their
corresponding DFT counterparts. Missing minima and transition state matches are
highlighted in red, revealing gaps in MLIP landscape coverage, while additional (unmatched)
stationary points are omitted. Most models generate minima structures which are close to their DFT counterparts, as seen by the overlapping small and large nodes. The materials-oriented MACE-MP-0b3 struggles to predict the correct structures for minima 1 and 5. We attribute this to the lack of molecule data in its training. The inclusion of landscape data in Allegro and
MACE training datasets produced marked improvements in stationary point
matching alongside reduced energy errors, as anticipated from previous work.\cite{CsanyiMW23} The systematic difficulty in
transition state discovery using MLIP energy landscapes emerges as a critical
limitation, with successful transition state searches proving challenging
across all tested architectures. 

\newpage
\subsubsection{Reproducing kinetics}

As a final test we computed mean first passage times (MFPT) between nodes within these networks, using the graph transformation procedure.\cite{TrygubenkoW06a, TrygubenkoW06b,SharpeW21c} The MFPT provides a physically meaningful and succinct metric for the overlap between DFT and MLIP KTNs, which includes all the pathways with appropriate weights, whether physical or not.
It also depends on the accuracy with which the DFT stationary points 
are reproduced.

\begin{figure}[tp!]
        \centering
        \includegraphics[width=1\linewidth]{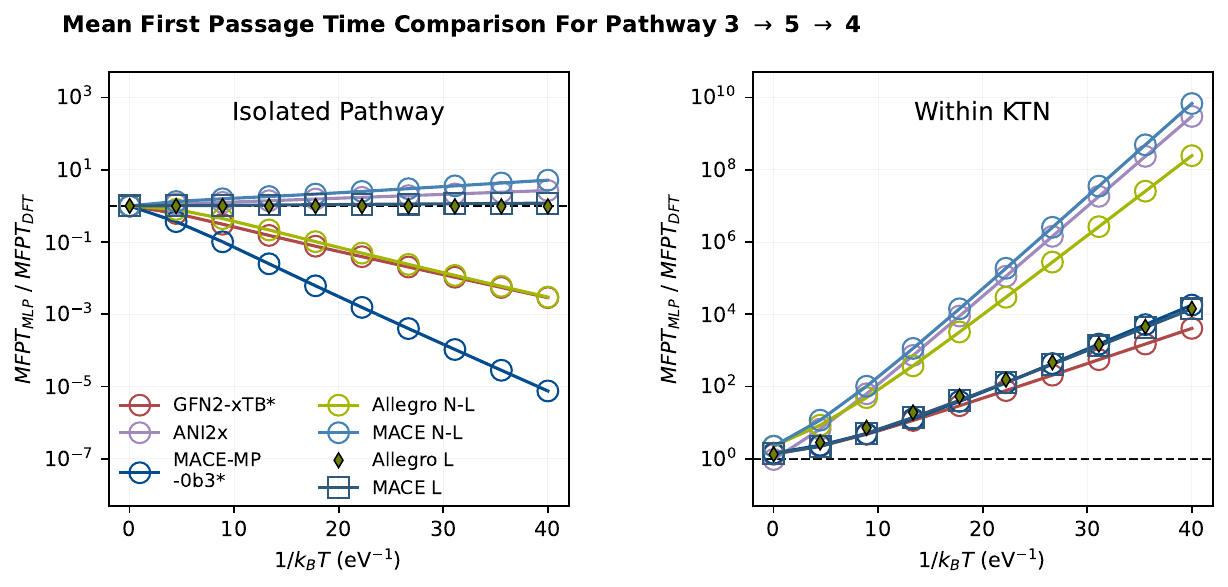}
        \caption{Mean first passage times between nodes 3 and 4 (c.f. Fig. \ref{fig:ktncomparison}) as computed from various MLIP KTNs and reported with respect to the DFT values. In the left panel we show the case where the path 3-5-4 is isolated from the rest of the network, while in the right panel we retain the entire KTN.}
        \label{fig:meanFirstPassageTimes}
\end{figure}

In Fig.~\ref{fig:meanFirstPassageTimes} we present MFPTs for a pathway that is
well described by most models (with the exception of NequIP), corresponding to
the nodes 3$\rightarrow$5$\rightarrow$4 (c.f. Fig.~\ref{fig:ktncomparison} and Fig.~\ref{fig:fig1}). We
compute MFPTs in the regime where an unbranched pathway is extracted from the 
KTN (left panel), and where the pathway is embedded in the rest of the network
(right panel). For ease of presentation we only show the best and worst results
for foundational models (therefore we disregard SO3LR and AIMNet2), and
concentrate on the comparison between N-L and L MACE and Allegro models.

L models perform significantly better than N-L counterparts, with the
low-temperature results for the pathway embedded within the KTN leading to a
factor of 10$^4$ and 10$^6$ improvement in the MFPT in the case of Allegro and
MACE, respectively. GFN2-xTB has a low-temperature offset of 10$^3$ in both the
embedded and unbranched single path case, while MACE-MP-0b3 leads to better MFPT results in the
embedded case.

\section{Discussion}

Machine learning potentials offer computational speed that facilitates longer
simulations and more extensive exploration of potential energy surfaces
compared to traditional quantum chemical methods. However, this improved
sampling capacity raises important questions about the ability of these models
to accurately reproduce the correct organization of the global energy landscape, and how
training data composition influences this ability. Kinetic transition networks
provide a concise description of the potential energy surface organization, and their
faithful reproduction is essential for computing reliable transition rates.
While comprehensive landscape searches remain largely computationally
intractable for density functional theory calculations, machine learning potentials make such
sampling feasible, creating an opportunity to rigorously test
their global accuracy.

We address this challenge by introducing benchmarks for KTN reproduction,
providing a framework to validate machine learning potential performance in
regions crucial for both thermodynamics and kinetics. We computed complete
reference KTNs for six molecules from the rMD17 dataset using hybrid DFT,
creating the Landscape17 dataset containing all minima and transition states
for ethanol, malonaldehyde, paracetamol, azobenzene, salicylic acid, and
aspirin. Using this benchmark, we systematically evaluated three categories of
models: foundational models (ANI2x, AIMNet2, MACE-MP-0b3, SO3LR),
system-specific models (Allegro, MACE, NequIP), and semi-empirical methods
(GFN2-xTB).

To isolate the effects of training data composition on landscape reproduction,
we compared two distinct approaches for system-specific models: training
exclusively on molecular dynamics data from rMD17 (N-L models) versus
augmenting MD data with approximate steepest-descent paths from transition states (L
models). This comparison enables us to validate performance beyond 
force and energy errors, while directly probing how kinetically relevant
training data affects potential energy surface representation.

The results reveal a systematic tendency for all machine learning potentials
to overestimate the number of stationary points:
MLIPs frequently generate unphysical conformations, despite basin-hopping steps
constrained to only dihedral rotations. This behavior reflects substantial
artifactual structure in machine learning potential energy surfaces compared to the DFT
reference, a phenomenon that can be quantitatively analyzed using the
landscape exploration methodology proposed here. It is noteworthy that the GFN2-xTB
framework exhibited markedly different behavior, with no unphysical conformations arising
during exploration.

The incorporation of landscape-specific data into system-specific model
training had two main effects: while increasing the total number of discovered
stationary points (including both physically meaningful and spurious
configurations), it simultaneously enhanced model performance across multiple
metrics. These augmented models exhibit superior accuracy in reproducing
DFT-calculated energies and forces at stationary points, while also generating
KTNs with improved rates and overall fidelity to reference DFT landscapes.
These results suggest that incorporating kinetically relevant training data can
effectively improve the representation of potential energy surface curvature in
machine learning models.

Nevertheless, achieving perfect correspondence with DFT reference data remains
elusive across all the models we have tested. This becomes particularly
pronounced for transition state prediction, where all the machine learning
potentials identified fewer than half the total number of transition states
present in DFT landscapes. The persistent challenges in transition state
identification highlight fundamental limitations in current machine learning
potential architectures and training methodologies, even for the set of simple,
canonical molecules that constitute the Landscape17 dataset.

Based on these results, we suggest that reproducing full conformational
landscapes of small molecules could serve as a useful benchmark for
evaluating future machine learning potentials. This assessment framework
provides a test of model fidelity beyond the usual energy and force
prediction metrics, directly probing the ability to capture the 
features that are essential for kinetic modelling and the tendency of MLIPs to
generate stable unphysical structures.

\section{Methods}

\subsection{DFT landscapes} \label{DFT_landscapes}

\subsubsection*{Energy landscape framework}

The energy landscape framework provides a comprehensive approach to mapping
surface topography through the identification and characterization of
stationary points.\cite{Wales18, JosephRCMW17,Wales18,RoederJHW19,Niroomand2024Review} These are atomic
configurations at which the gradient vanishes and we focus on
local minima and the transition states that connect them, which are distinguished by their Hessian eigenvalue spectrum. Local
minima exhibit only positive and zero Hessian eigenvalues, indicating that any
displacement of internal coordinates increases the energy. Transition states are defined as first-order
saddle points with exactly one negative eigenvalue, corresponding to a local maximum
along the reaction coordinate, with positive curvature in the orthogonal
eigendirections (aside from the zero eigenvalues corresponding to overall rotation and translation).\cite{Murrell1968}

These stationary points can be represented by weighted graphs, known as kinetic
transition networks (KTNs),\cite{NoeF08,pradag09,Wales10a} where minima serve as nodes and edges connect
minima that are directly linked by transition states.
Appropriate post-processing using standard tools of statistical mechanics and unimolecular rate theory
enables efficient computation of observable
thermodynamic and kinetic properties within well-defined approximations.\cite{Swinburne2020} 
In particular, the explicit inclusion of transition states,
which are more difficult to characterize using standard molecular dynamics, allows for
assessment of global kinetics and comparison of MLIP landscapes with the DFT reference.\cite{SharpeW21c,Woods2024}

\subsubsection*{Density functional theory calculations}

The reference potential energy landscapes were computed using density functional
theory with the $\omega$B97x hybrid-energy exchange correlation functional and a
6-31G(d) basis set within Psi4.\cite{Psi4} These settings are consistent with
the ones used to generate the ANI2x training data.\cite{ani2x} We applied
tight energy and density convergence criteria (\footnotesize
\textsc{E\_CONVERGENCE} \normalsize and \footnotesize
\textsc{D\_CONVERGENCE}\normalsize) of $10^{-9}$ Hartree and $10^{-9}$ a.u.,
along with extremely fine integration grids (100 radial and 770 spherical
points) and a restricted Kohn-Sham reference.

We validated the convergence, grid, and spin settings against published data
from rMD17, using the appropriate functional and basis set: PBE/def2-SVP. We
achieved energies and forces within 0.1\,meV/atom and 5\,meV/Å respectively,
well within the standard acceptable resolution of 1\,meV/atom and 10\,meV/Å.

For benchmarking energies of AIMNet2, MACE-MP-0b3 and SO3LR we recomputed the
DFT KTN stationary points with the same functionals (and dispersion correction)
as detailed in the original publications, specifically: $\omega$B97m-D3BJ, PBE
and PBE0-D3BJ. Following AIMNet2, we used the localized def2-TZVPP basis set
for all three cases, although we note that the DFT calculations for the
MACE-MP-0b3 and SO3LR training datasets originally used plane wave basis sets.
While using different functionals might yield different stationary points compared to
our $\omega$B97x calculations, we expect the results to be very similar for these small
molecules, as confirmed by our comparison results (Fig.~\ref{fig:comparison}).
Future work could include reconverging the stationary points using these
alternative functionals.

\subsubsection*{Minima identification}

Local minima were first collected from the basin-hopping global optimization
runs,\cite{lis87,Wales1997,waless99} an approach that has has been employed successfully 
for diverse molecular and abstract landscapes.\cite{Dicks2022} 
The basin-hopping exploration employed 
random angular perturbations
applied to flexible dihedral angles, as identified by the Atomic Simulation
Environment package.\cite{ase-paper} These surveys used 100 basin-hopping
steps with an accept/reject Metropolis condition equivalent to a temperature of 100\,K. The convergence criteria required either
the maximum force component to fall below $10^{-3}$\,eV/Å or the relative
energy change between steps to drop below $10^7$\,eV multiplied by machine precision.

\subsubsection*{Transition state location}

Transition state searches were performed between each minimum and its three
nearest neighbors, determined from the Euclidean distance with optimal alignment
via the \textsc{MINPERMDIST} routine.\cite{WalesC12,Griffiths2017} 
This procedure minimises the distance with respect to translation,
rotation, permutation and, additionally, we include the inversion operation.
The alignment is not deterministic when permutations are included, and employs
a shortest augmenting path algorithm\cite{jonkerv87} inside an iterative loop.\cite{WalesC12}
Distinct stationary points were distinguished by a root mean square distance threshold greater than
0.3\,Å and energy differences exceeding $10^{-3}$\,eV. 
Only one representative permutation-inversion isomer was retained for each
minimum and transition state.

Transition states were located using a two-step protocol starting with a 
nudged elastic band (NEB)\cite{JonssonMJ98,HenkelmanUJ00,HenkelmanJ00} calculation.
Initial pathways between
minima were generated through linear interpolation in internal coordinates
after endpoint alignment. 
The NEB algorithm optimizes these interpolations of
20 images with a 50\,eV/Å spring constant and convergence criterion of
$10^{-2}$\,eV/Å for the maximum force component.
This double-ended phase of the calculation only needs to converge sufficiently to
identify the local maxima in the profile,
which are taken as starting points for accurate refinement using
hybrid eigenvector-following.\cite{munrow99,KumedaMW01}
Here the smallest non-zero eigenvalue and the corresponding eigenvector are
obtained using a variational method, with minimisation in all orthogonal directions.
Convergence required RMS forces below $3 \times 10^{-2}$\,eV/Å.
Transition states were verified through Hessian analysis, confirming exactly one
negative eigenvalue. 
Only transition states with barriers below 1\,eV were retained. 
We do not assume that a path links the two original end minima from the NEB phase, even when there is
only a single transition state in the profile.
The connectivity of each transition state is always established from the corresponding pathways.
In general, there may be multiple transition states between two minima, and there may be gaps in the
connection profile.
The two-phase procedure is applied until a complete discrete path is obtained, using the
missing connection algorithm to propose new pairs of minima for additional searches.\cite{CarrTW05}

\subsubsection*{Pathways and network construction}

Each transition state connects two local minima, identified through
minimization from perturbed transition state geometries.
Perturbations of approximately $0.3\mathbf{y}$ ($0.6\mathbf{y}$ for flatter
modes, such as the cis-cis transition state in azobenzene) were applied along
parallel and antiparallel to the normalized eigenvector, $\mathbf{y}$, corresponding to the negative eigenvalue, followed by LBFGS minimization.
All intermediate configurations during minimization were
stored to map the approximate steepest-descent paths from transition states to
connected minima, excluding initial configurations with force components exceeding 2\,eV/Å.

\subsubsection*{Dataset construction}

Complete KTNs were constructed for six molecules from the rMD17
dataset:\cite{MD17, rMD17} ethanol, malonaldehyde, salicylic acid, azobenzene,
paracetamol, and aspirin. These molecules were selected as they contain
multiple distinct isomers, enabling meaningful landscape
analysis. All calculations utilized the TopSearch Python
package,\cite{topsearch} (or OPTIM program\cite{OPTIM}, for azobenzene, see Appendix \ref{sec:appendix-landscape17}). The resulting compilation of minima, transition
states, and approximate steepest-descent pathways constitutes the Landscape17 dataset.
Furthermore, we include the Hessian eigenspectrum at the stationary points for
reference.

\subsubsection*{Mean first passage times}

Mean first passage times were computed from the KTNs using the graph transformation
approach,\cite{TrygubenkoW06a, TrygubenkoW06b,SharpeW21c} as implemented in the PyGT
package.\cite{swinburne_defining_2020} 
The individual rates for each elementary transition between
minima directly connected by a transition state were computed using harmonic transition state theory.\cite{Eyring35,EvansP35,forst73,Laidler87}
The same choice is made consistently to compare the results from the MLIP landscapes and the reference
DFT KTN.
More accurate elementary transition rates could be used instead, including methods
based on explicit dynamics and associated path sampling.\cite{BolhuisCDG02,DellagoB09,FaradjianE04,AllenFW06b,VanErpMB03,VanErpB05}
However, the harmonic transition state theory approach is a common choice for studies that
treat the global dynamics of a KTN,\cite{SorensenV00,:/content/aip/journal/jcp/141/16/10.1063/1.4898664,Sharia_2016,SwinburneP18}
and is appropriate for the comparison that we require here.
We compute the rates between the states corresponding to the
nodes 3-4 in Fig.~\ref{fig:ktncomparison}. For the unbranched path, we retain only
nodes 3, 5 and 4 in the network, removing direct connections between nodes 3-4.

\subsection{Training MLIPs}

We trained and evaluated a range of machine learning potentials:
Nequip,\cite{Nequip} MACE,\cite{MACE} Allegro,\cite{Allegro} ANI2x,\cite{ani2x}
AIMNet2,\cite{Aimnet2} MACE-MP-0b3 (medium),\cite{mace-mp-0b3}
SO3LR;\cite{kabylda_molecular_2025} but also a semi-empirical method:
GFN2-xTB\cite{bannwarth_gfn2-xtbaccurate_2019} (via
DFTB+).\cite{hourahine_dftb_2020} For detailed architecture descriptions, we
refer readers to the original publications. Among these, ANI2x, AIMNet2,
MACE-MP-0b3, SO3LR, and GFN2-xTB are transferable potentials that can be used
without retraining, while MACE, Allegro, and NequIP are architectures requiring
training for each application. We conducted ANI2x and AIMNet2 evaluations on
single CPUs, with all other models were evaluated on single NVIDIA A100 GPUs.

We train models both with and without the incorporation of landscape data. In
both cases, the custom models were trained from scratch using the
hyperparameters specified in their respective publications. For the
non-landscape (rMD17) dataset, we used the same structures in a 950/50
test/train split, maintaining consistency with the original papers. For the
landscape models we additionally included 40\% of the pathway data
from Landscape17 for the corresponding molecules (up to 500 structures total),
reserving 10\% for validation. The remaining 60\% of pathway configurations (up
to 1000 structures) served as additional test data. For ethanol and malonaldehyde we add fewer than 40\% data points, in line with their low number of atoms. Training details and model
performance metrics are provided in Appendices \ref{sec:appendix-landscape17}
and \ref{app:training}.

\subsection{MLIP landscapes}

We generated MLIP landscapes using the methodology described in
Sec.~\ref{DFT_landscapes}. For each MLIP, we created the final KTN by
initiating the landscape generation process from 20 different starting
structures taken from rMD17. Each starting point produced a separate KTN, which
we then merged while eliminating permutation-inversion isomers according to the same
similarity criteria used for the reference landscapes. The Appendix contains a
flowchart illustrating this process (Fig.~\ref{fig:flowchartLandscapeGen}).

\subsection{Landscape analysis}

We evaluated the quality of the MLIP landscapes with respect to the DFT references
using various metrics detailed below and in Appendix
\ref{appendix:mlpLandscapes}, using the methodology outlined in
Figs.~\ref{fig:flowchartOverview}, \ref{fig:flowchartSubmodules} and
\ref{fig:flowchartcomparisonAlgorithm}.

\subsubsection*{Unphysical structures}

A significant limitation of current MLIPs is their tendency to produce
unphysical molecular structures. To identify these unphysical structures,
we compare each MLIP stationary point against the initial bonding framework.
Changes in the adjacency matrix indicate inappropriate bond formation or
breaking, which should not occur under the simple dihedral angle rotations that we
implement. We detect these issues by monitoring changes in SMILES
representations (via RDKit\cite{rdkit}) and hydrogen atom coordination numbers to identify spurious
multiple bonding.

\subsubsection*{Landscape comparison}

As described above, application of the energy landscape framework to a given
potential energy surface produces a KTN.\cite{NoeF08,pradag09,Wales10a}
Comparing these networks allows us to evaluate the similarity between potential
energy surfaces from different methods, particularly in both low-energy
(minima) and higher-energy (transition state) regions. Here, we describe our
methodology for quantifying similarities between two networks: a reference
DFT-generated landscape and one produced by an MLIP.

Our comparison process begins by removing unphysical structures from the MLIP
networks and checking for duplicate structures. We then conduct the comparison
along two tracks, as mentioned in the Results section. The first track
identifies \textit{exact matches} between DFT and MLIP minima and transition
states, while the second track computes similarities between the
\textit{closest pairs} of minima using less stringent matching criteria.

Both approaches start by computing similarity matrices between minima in both
networks using the \textsc{MINPERMDIST} routine\cite{WalesC12,Griffiths2017} to compute RMSD
values. We then match minima using a greedy algorithm that pairs each DFT
minimum with its closest MLIP counterpart (we note that Flowchart
\ref{fig:flowchartcomparisonAlgorithm} shows an easier computational
alternative for finding exact matches that is in fact equivalent to computing
the similarity matrix). For exact matching, we add the requirement that matches
must be within 0.3\,Å RMSD. Successful matches meet this criterion; failed
matches do not. The number of additional minima (shown in Fig.~\ref{fig:comparison}
\textbf{a}) is calculated by subtracting successful matches from the total MLIP
minima count. 

For transition state comparisons, the closest-match approach uses the same
algorithm as for minima. The exact-match approach is more stringent, requiring
both a 0.3\,Å RMSD threshold and that the transition states connect minima that
are themselves exactly matched between DFT and MLIP networks. These results
appear in Fig.~\ref{fig:comparison} \textbf{b}.

In almost all cases, MLIP networks contain more minima than the DFT counterparts,
resulting in consistent closest match counts across models for each molecule
(equal to the DFT minima count). This observation enables meaningful comparison of average
RMSD and energy RMSE across models (Fig. \ref{fig:comparison} \textbf{c} and
\textbf{d}). However, for transition states the counts vary significantly
between models and there are often fewer MLIP than DFT transition states (Fig.
\ref{fig:mlpLandscapes} \textbf{b}), preventing reliable cross-model
comparison.

\section*{Author Contributions}

LD, FT and EPK conceived the project. LD, VC, JM and DJW generated
the DFT reference data. VC and FT trained the MLIPs. VC generated the MLIP landscapes. VC, FT, LD, JM and EPK performed
the data analysis. VC, LD and FT wrote the original draft. All authors read,
edited and approved the final manuscript.

\section*{Acknowledgments}

This work was performed using resources provided by the Cambridge Service for
Data Driven Discovery 
(www.csd3.cam.ac.uk), provided by Dell EMC and Intel using
Tier-2 funding from the Engineering and Physical Sciences Research Council
(capital grant EP/T022159/1), and DiRAC funding from the Science and Technology
Facilities Council (www.dirac.ac.uk). VC acknowledges the computational
resources obtained through the University of Cambridge EPSRC Core Equipment
Award (EP/X034712/1) and EPSRC IAA award number G116766.

We thank Gábor Csányi for useful discussions.
\section*{Competing interests}

All authors declare no financial or non-financial competing interests.

\section*{Code availability}

All DFT data described in this work was generated using the TopSearch Python
package (v0.0.3),\cite{topsearch} available at
\url{https://github.com/IBM/topography-searcher}. The MLIP and GFN2-xTB
landscapes were produced produced using the forked branch available at: \url{https://github.com/VladCarare/topography-searcher/tree/analysis}, while the analysis employed the forked branch:
\url{https://github.com/VladCarare/topography-searcher/tree/mlp_run}. An example of how to use the generation and analysis code is available at \url{https://github.com/VladCarare/mlp-landscapes/tree/main}.

\section*{Data availability}
The Landscape17 dataset, which was generated and analyzed during the current study, is available in the Figshare repository at DOI: \url{https://doi.org/10.6084/m9.figshare.29949230}.\cite{landscape17dataset}

\bibliographystyle{naturemag}
\bibliography{bibliography}

\newpage
\appendix
\renewcommand{\thesubsection}{\Alph{subsection}}
\renewcommand\thefigure{A.\arabic{figure}}
\renewcommand\thetable{A.\Roman{table}}
\setcounter{figure}{0}

\section{Landscape17}
\label{sec:appendix-landscape17}
The Landscape17 dataset comprises the KTNs for the molecules within rMD17
that have multiple distinct (excluding permutational-inversion isomers) structures: ethanol,
malonaldehyde, salicylic acid, paracetamol, azobenzene and aspirin. The data
compiled for each molecule contains the atomic configurations, energies, and
forces for all minima, transition states, and every configuration on the
calculated paths from transition states to their connected minima. 
We describe each KTN and provide technical validation of the dataset in the following subsections.

\subsection*{Landscapes}

\begin{figure}[h!]
    \centering
    \includegraphics[width=0.75\linewidth]{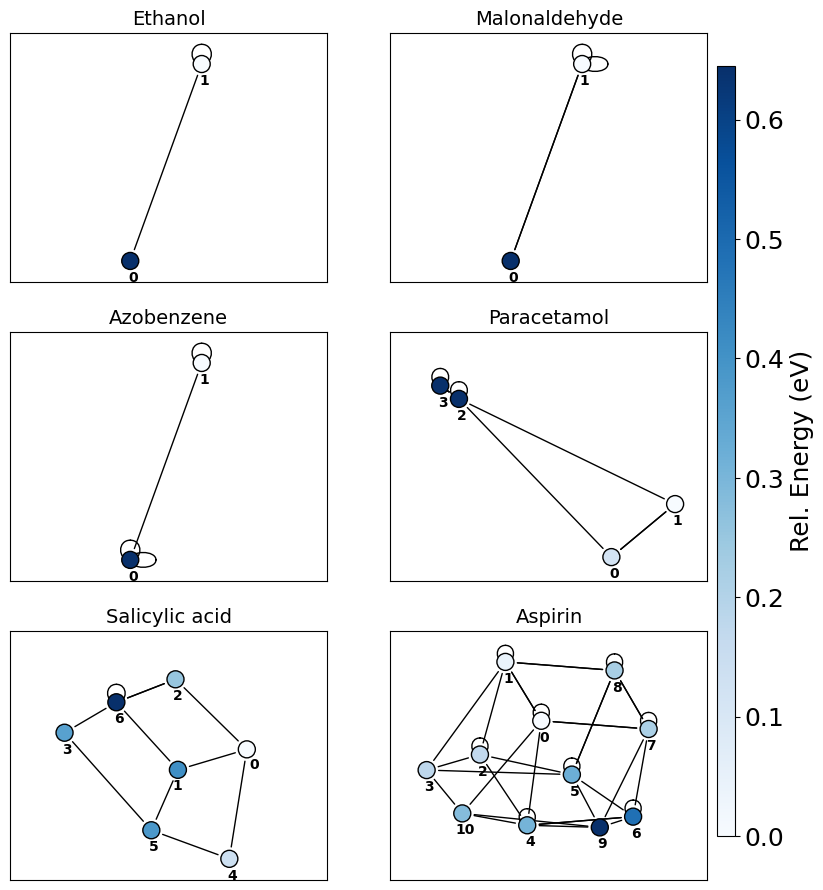}
    \caption{Visual depiction of the KTN for each molecule. Each node is a
minimum on the potential energy surface, and edges are drawn when there is a
transition state that directly connects two minima. Self-loops indicate
transition states that connect permutation-inversion isomers.}
    \label{fig:dft_ktn}
\end{figure}

We provide a brief visual overview of the set of KTNs in Fig.~\ref{fig:dft_ktn}.

\textbf{Ethanol} -- 
Ethanol has two distinct structures, which are related by
rotation around the CO bond. There are three stable orientations of the OH group
separated by $120^\circ$, and two of these are related by the inversion
operation.

\textbf{Malonaldehyde} -- Following rMD17 we consider only the keto form of
malonaldehyde. This tautomer has two distinct minima for our choice of
DFT functional and basis set, and there are additional conformations
related by the inversion operation. The transition states arise largely from
rotation of one of the CC bonds, with the other fixed.

\textbf{Salicylic acid} -- Salicylic acid contains three rotatable bonds, each
with two states permitted by aromaticity, which correspond to rotation by
$180^\circ$. One of these possible conformations results in clashing hydrogens,
leaving seven distinct conformations.\cite{Aarset2006}

\textbf{Paracetamol} -- The potential energy surface of paracetamol contains
four distinct local minima.\cite{Varela2013} The peptide bond can be either cis or trans
and the OH can be rotated by $180^{\circ}$.

\textbf{Aspirin} -- Previous studies have highlighted nine local minima of
aspirin,\cite{Glaser2001} of which two contain the majority of the
equilibrium population at room temperature.\cite{Saadeh2023} We also found two additional local minima.
Naively assuming only two available rotational states for the two bonds of the
carboxylic and ketone groups gives 16 possible conformations. Rotation of the
whole ketone group results in conformations related by the inversion operation,
leaving eight distinct conformations. In some configurations rotation of the
carboxylic acid group can lead to more than two stable rotational states,
resulting in the final set of eleven minima.

\textbf{Azobenzene} -- Azobenzene has two distinct minima related by rotation
of the central non-aromatic bonds; these conformations can be considered cis
and trans isomers. The exact nature of the transition states depends upon the
electronic state,\cite{Cembran2004} and here we choose the ground state, ignoring
conical intersections. These two distinct minima exhibit two enantiomeric
conformations each, which are connected by a single transition state in the
case of the trans conformation, and by two energetically-similar transition
states in the case of the cis conformation. A large barrier separates
the trans-trans transition state from one of the cis-cis transition
states.\cite{azobenzene_complete} Two of the pathways exhibit
branch points, where the Hessian eigenvalue associated with a perpendicular mode
passes through zero. The characterization of these pathways required special attention, and here we used the OPTIM program\cite{OPTIM} from the Cambridge Energy Landscapes software suite (\url{https://www-wales.ch.cam.ac.uk/OPTIM}). To the best of our knowledge, these paths have not been described before.

\subsection*{Technical Validation}

There are several components of the landscapes that should be checked: the
validity of the stationary points, their correct assignment as minima or
transition states, and the enumeration of the relevant stationary points from
the PES. It is straightforward to validate stationary points and their
character, as described in the following subsections, but it is more challenging to
validate the enumeration of all relevant minima and transition states. However,
we find minima in line with previous studies and expectations from
enumeration of rotatable bonds. Searches for transition states between these minima were guided by
the separation in Euclidean distance, ensuring that the relevant transition states
are located.

\subsubsection*{Stationary points}

Validating molecular conformations as stationary points requires a single-point
force calculation. We extract each conformation from the KTNs and recompute the
forces to ensure they satisfy our convergence criteria. For minima we specified
a maximum force component below $1 \times 10^{-3}$\,eV\,\AA$^{-1}$, with
limited exceptions due to minimisation terminating from the energy change
rather than force. In all cases there are no force components greater than $3
\times 10^{-3}$\,eV\,\AA$^{-1}$. For transition states, which are more
difficult to locate, the convergence criterion was that the maximum force component must be
less than $3 \times 10^{-2}$\,eV\,\AA$^{-1}$. We checked that this convergence
criterion is met for all transition states and the corresponding gradients are
stored within the data records.

\subsubsection*{Minimum or transition state}

We assign each stationary point as a minimum or transition
state from the Hessian eigenspectrum. Minima should have only non-negative eigenvalues,
while transition states have a single negative
eigenvalue.\cite{Murrell1968} We extracted the coordinates for each stationary point and computed
the Hessian matrix eigenvalues, which are stored within the data
records. Note that the eigenvalues include six zeros corresponding to
overall translation and rotation. These eigenvalues are the smallest in
magnitude in all cases and are well separated from the vibrational modes,
confirming tight convergence of the geometry optimisation.

\clearpage
\newpage
\section{Data distributions} 
\label{sec:appendix-datadistr}

In Table \ref{tab:molecule_data_sd_md_splits} we present the number of
approximate steepest-descent path samples (SD) along with the training/validation/testing
splits for the Non-Landscape and Landscape datasets (where the Landscape models
are trained using the combination of N-L and L datasets).
    
        \begin{table}[htbp]
            \centering
	    \caption{Molecular dataset statistics after removing additional
permutation-inversion isomers. We note that L
models are trained on the combination of N-L and L datasets.}
            \label{tab:molecule_data_sd_md_splits}
            \small 
            \begin{tabular}{@{}l*{9}{c}@{}} 
            \toprule
            Molecule & \begin{tabular}[c]{@{}c@{}}DFT\\ \# min\end{tabular} & \begin{tabular}[c]{@{}c@{}}DFT\\ \# TS\end{tabular} & \begin{tabular}[c]{@{}c@{}}DFT\\ \# SD\end{tabular} & \begin{tabular}[c]{@{}c@{}}N-L\\train\end{tabular} & \begin{tabular}[c]{@{}c@{}}N-L\\val\end{tabular} & \begin{tabular}[c]{@{}c@{}}N-L\\test\end{tabular} & \begin{tabular}[c]{@{}c@{}}L (only)\\train\end{tabular} & \begin{tabular}[c]{@{}c@{}}L (only)\\val\end{tabular} & \begin{tabular}[c]{@{}c@{}}L (only)\\test\end{tabular} \\
            \midrule
            ethanol & 2 & 2 & 123 & 950 & 50 & 1000 & 20 & 2 & 101 \\
            malonaldehyde & 2 & 4 & 414 & 950 & 50 & 1000 & 112 & 12 & 290 \\
            salicylic & 7 & 11 & 1215 & 950 & 50 & 1000 & 438 & 48 & 729 \\
            azobenzene & 2 & 4 & 608 & 950 & 50 & 1000 & 219 & 24 & 365 \\
            paracetamol & 4 & 9 & 1869 & 950 & 50 & 1000 & 450 & 50 & 1000 \\
            aspirin & 11 &  37 & 6944 & 950 & 50 & 1000 & 450 & 50 & 1000 \\
            \bottomrule
            \end{tabular}
        \end{table}

\begin{figure}[htp]
    \centering
    \includegraphics[width=0.89\linewidth]{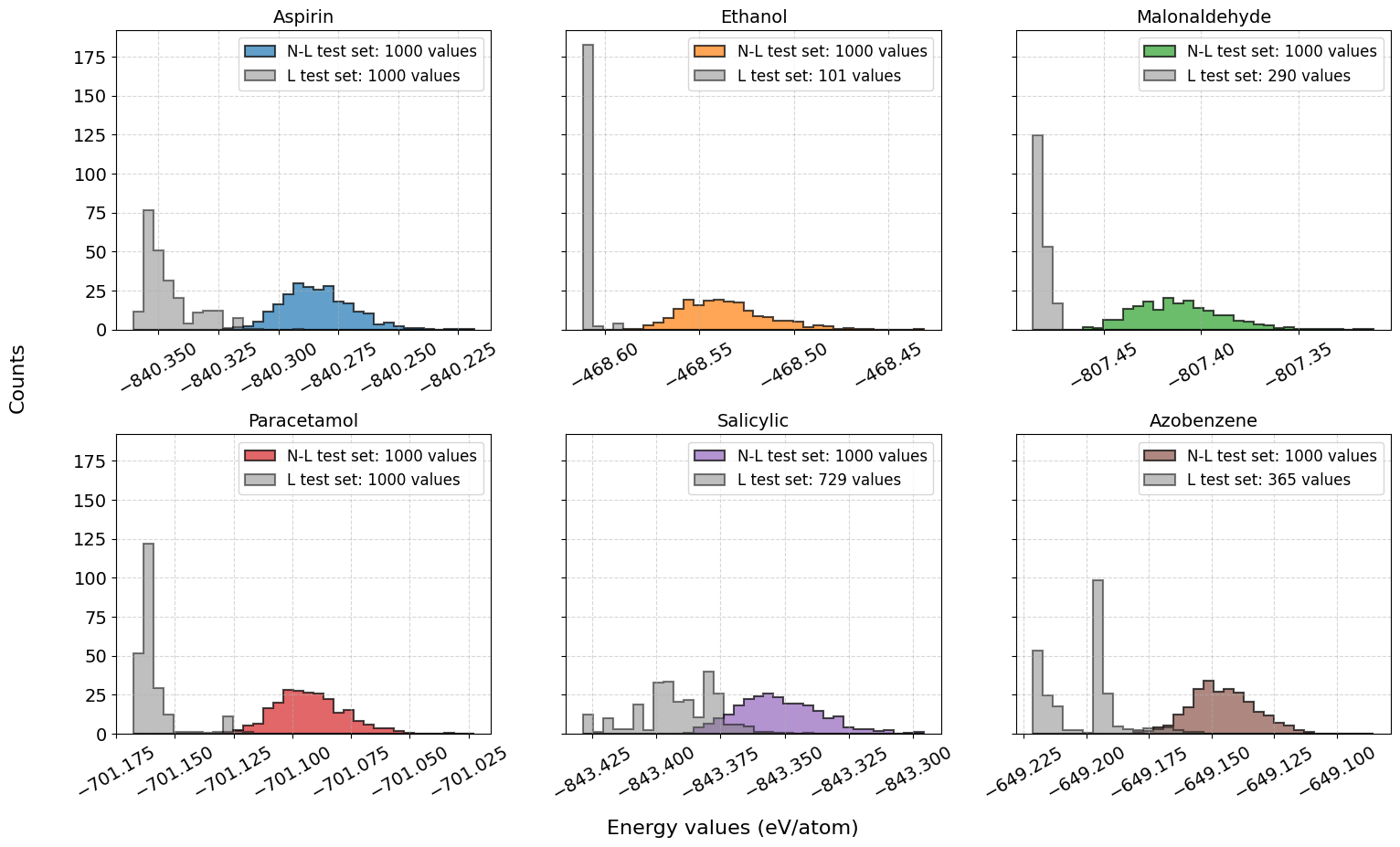}
    \caption{Energy distributions for both non-landscape and landscape test sets, generated using molecular dynamics and 
pathway configurations, respectively.}
    \label{fig:mdsdenergydistribution}
\end{figure}

\begin{figure}[htp]
    \centering
    \includegraphics[width=0.89\linewidth]{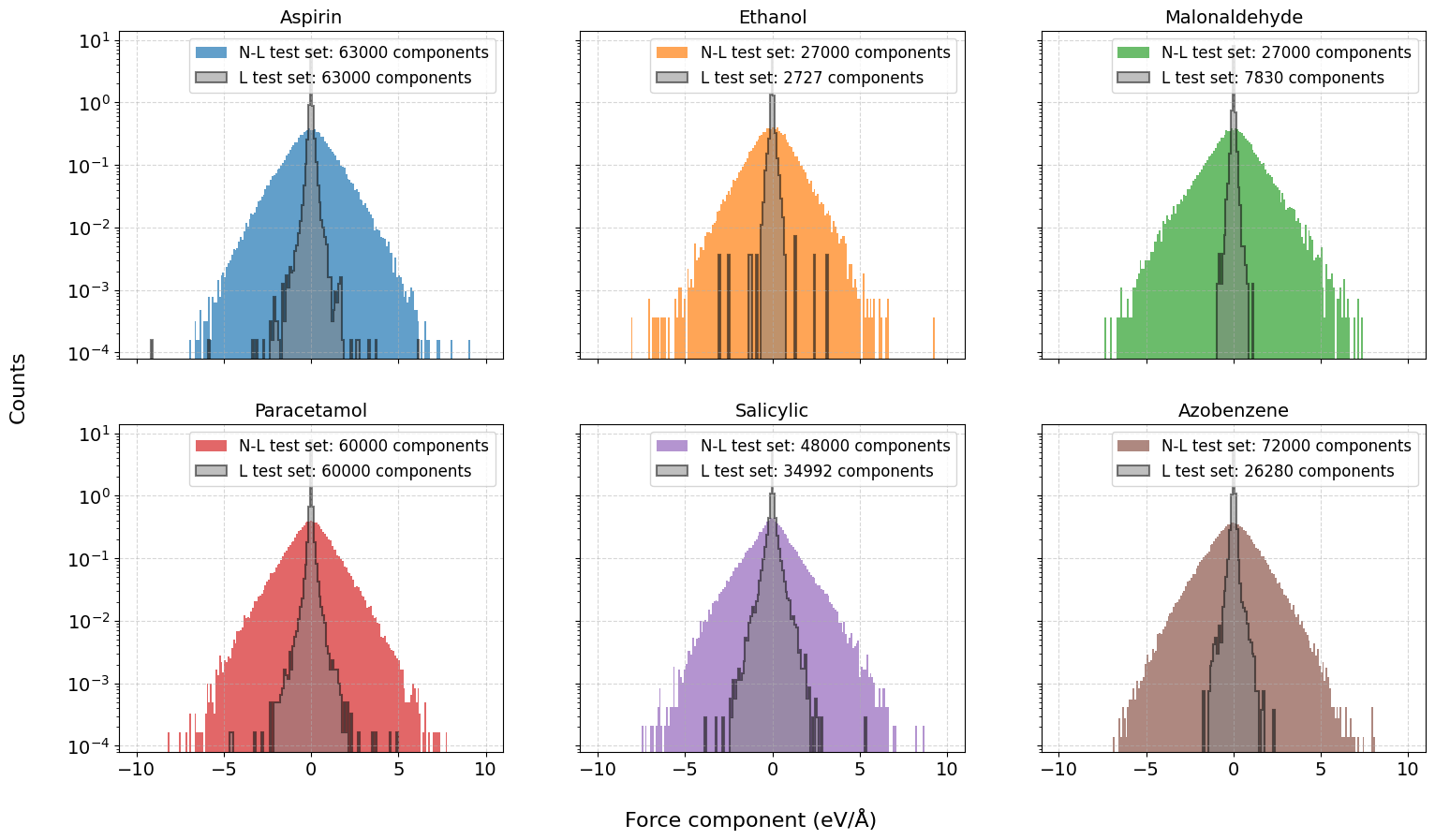}
    \caption{Force distributions for both non-landscape and landscape test sets, generated using molecular dynamics and 
pathway configurations, respectively.}
    \label{fig:mdsdforcedistribution}
\end{figure}

In Figs.~\ref{fig:mdsdenergydistribution} and \ref{fig:mdsdforcedistribution}
we show the energy and forces histograms for the N-L and L test sets, to
complement Fig. \ref{fig:fig1} \textbf{B} shown in the Results section.

\clearpage 

\begin{figure}[htp]
    \centering
    \includegraphics[width=0.9\linewidth]{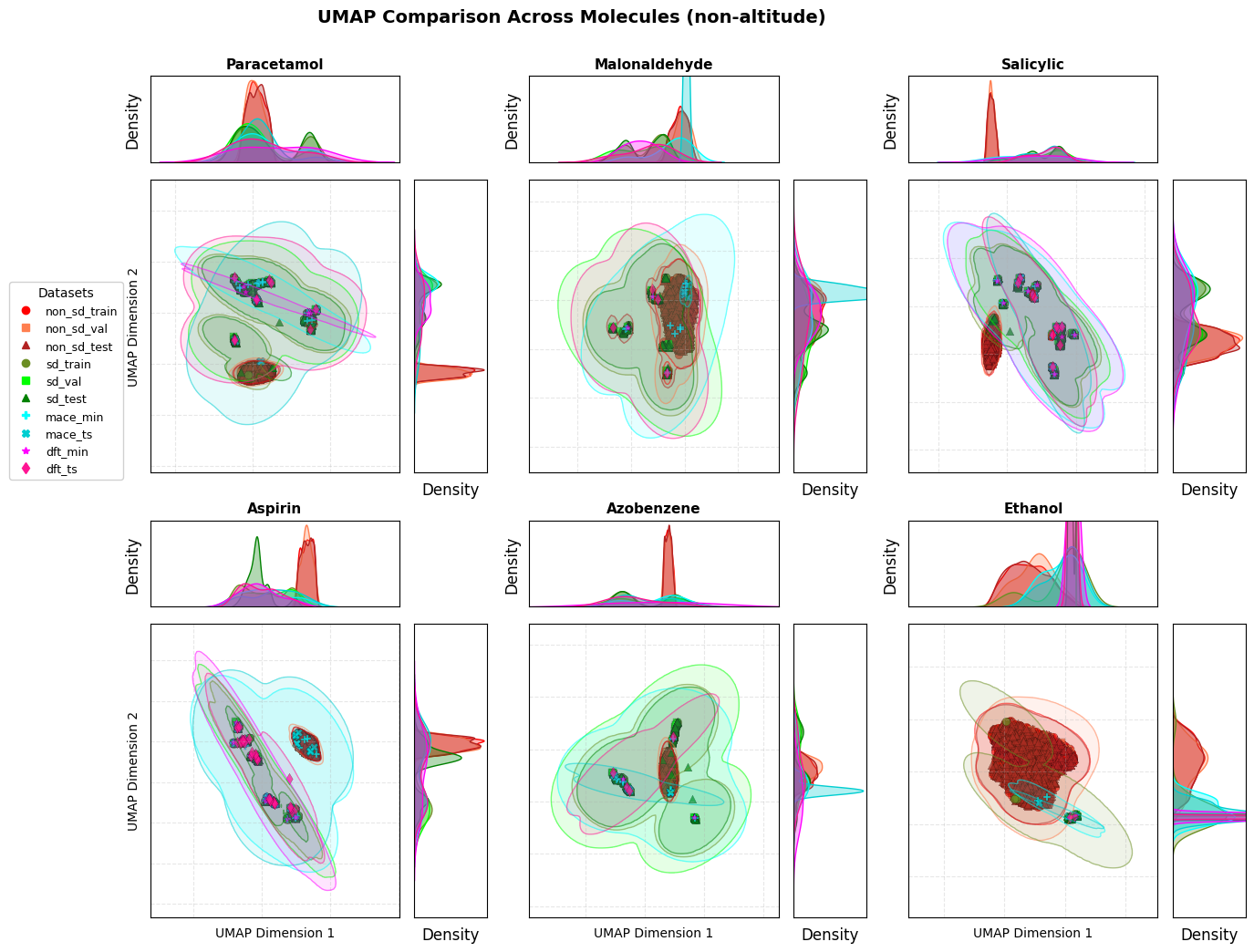}
    \caption{UMAP projections of subsets of data: Non-Landscape (here written
as \textit{non\_sd}), Landscape (\textit{sd}) and stationary points of DFT and
MACE KTNs. The descriptors used for the projections are the invariant
embeddings of the corresponding data, as given by the Non-Landscape model of
the respective molecule. The density and contour lines are computed using
kernel density estimation within \textsc{SciPy}.}
    \label{fig:umapmd}
\end{figure}

\begin{figure}[htp]
    \centering
    \includegraphics[width=0.9\linewidth]{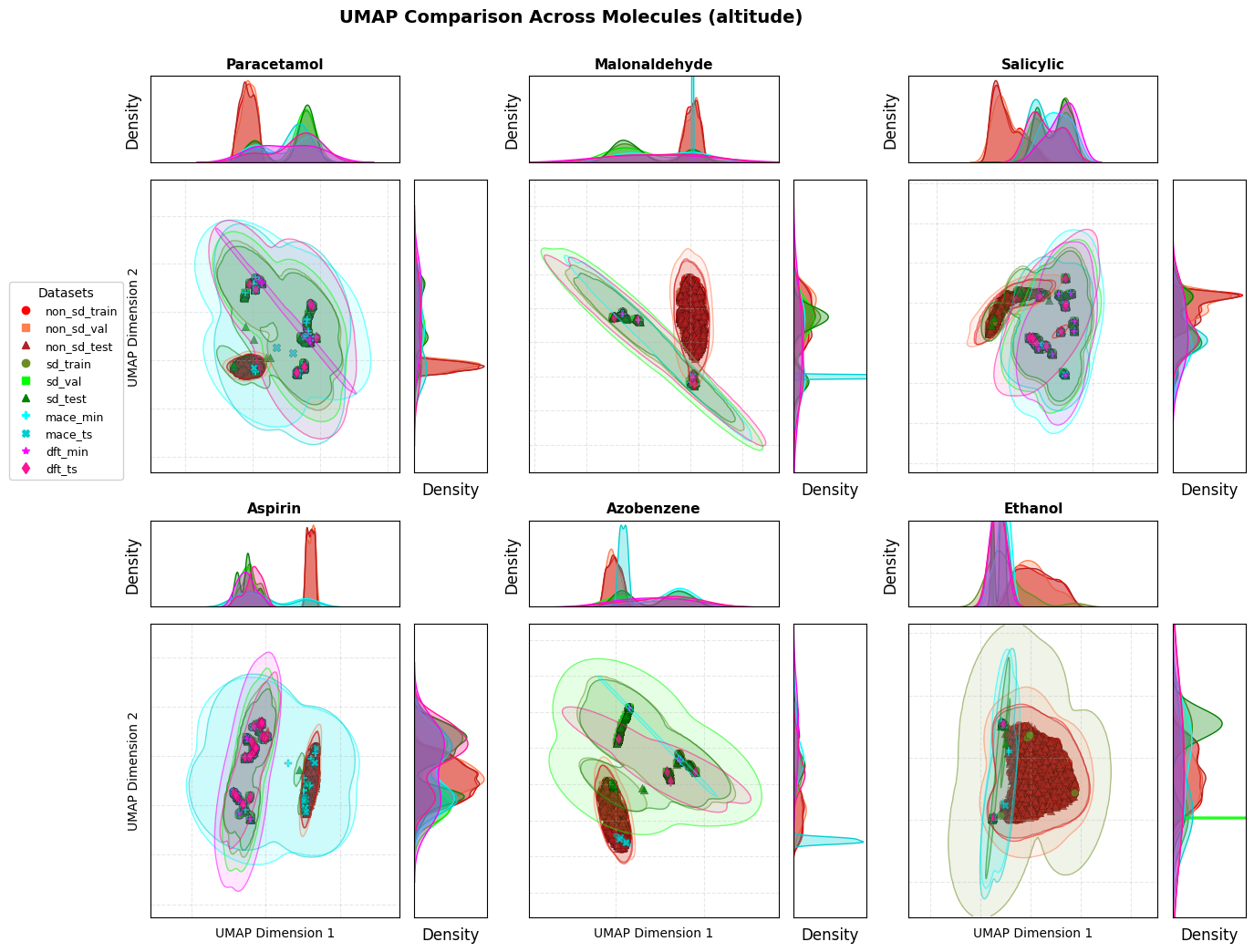}
    \caption{UMAP projections of subsets of data: Non-Landscape (here written
as \textit{non\_sd}), Landscape (\textit{sd}) and stationary points of DFT and
MACE KTNs. The descriptors used for the projections are the invariant
embeddings of the corresponding data, as given by the Landscape model (c.f.
Fig. \ref{fig:umapmd}) of the respective molecule. The density and contour
lines are computed using kernel density estimation within \textsc{SciPy}.}
    \label{fig:umapsd}
\end{figure}

Figures \ref{fig:umapmd} and \ref{fig:umapsd} (extending Fig. \ref{fig:fig1}
\textbf{C}) show the UMAP projections of the Non-Landscape and Landscape
datasets along with the stationary points of the DFT and MACE KTNs, on the UMAP
axes given by the MACE descriptors of the training set of the N-L model
(Fig.~\ref{fig:umapmd}) and L model (Fig.~\ref{fig:umapsd}), respectively. We have used standard hyperparameters for the UMAP: a dist of 0.1 and 15 neighbors.

\clearpage
\newpage
\section{Training and validation of machine learning potentials} \label{app:training}

In the absence of published NequIP, Allegro and MACE potentials for rMD17, we
performed in-house training and compared the reported MAE values, tested on
100,000 structures, with our results tested on 1,000 structures, in Table
\ref{table:MAE_and_RMSE_for_non_altitude_training}.  We also provide the error
values for L models - those trained on Non-Landscape (rMD17) and Landscape data
- in Table \ref{table:MAE/_and_RMSE_for_altitude_training}). 
Tables \ref{table:MAE_and_RMSE_for_non_altitude_training_sd_testing} and
\ref{table:MAE_and_RMSE_for_altitude_training_sd_testing} show the errors for
the N-L and L models respectively, on the Landscape test sets.

\begin{figure}[htp]
    \centering
    \includegraphics[width=0.75\linewidth]{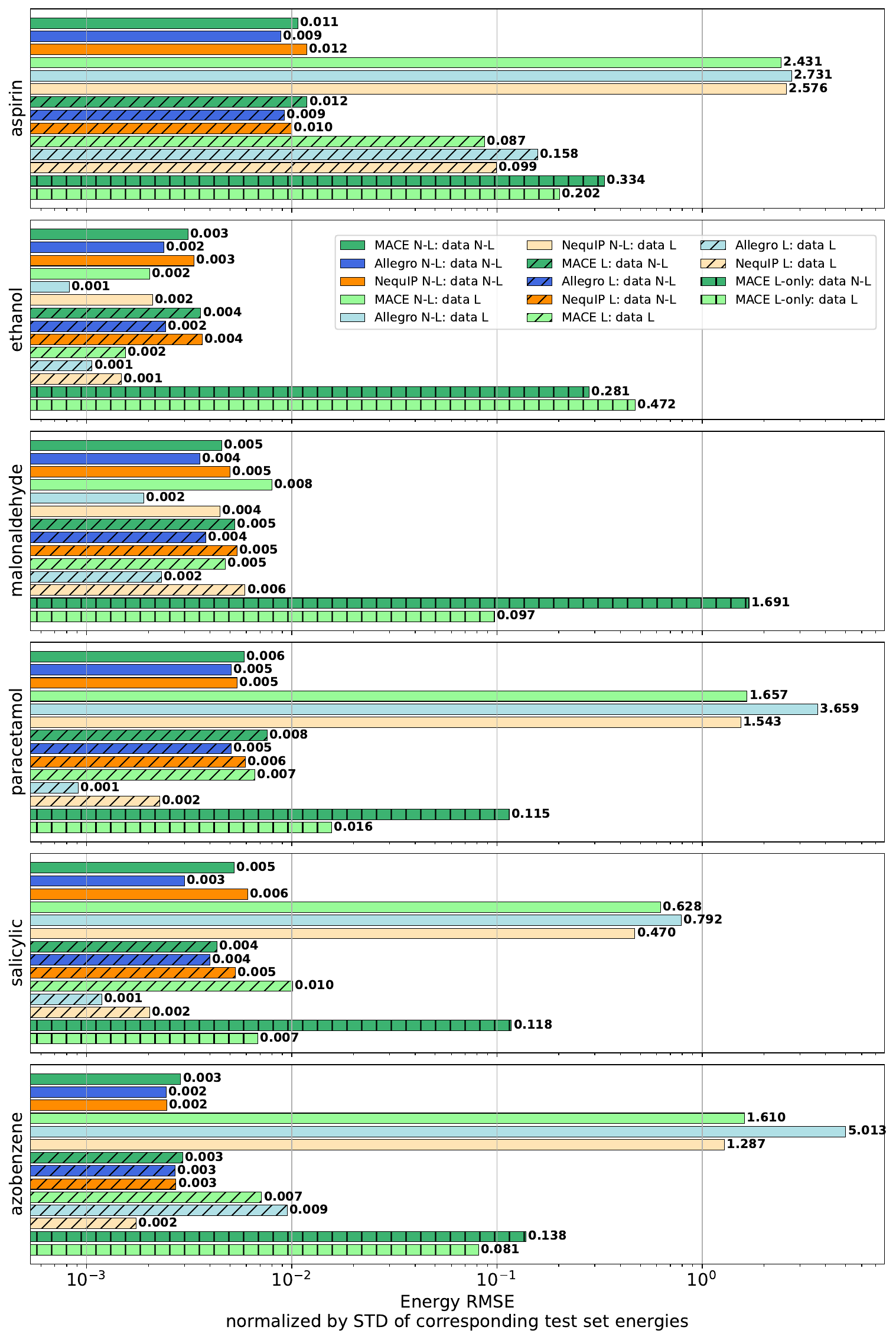}
    \caption{Energy RMSEs normalized by the standard deviation of the target
distribution. The legend follows the format: [Architecture] [N-L or L] : data
[N-L or L]; where the architecture is MACE, Allegro or NequIP, N-L or L refers
to the training dataset being composed of N-L or N-L and L data. The final
`data [N-L or L]' refers to the testing set over which RMSE is calculated and
normalized.}
    \label{fig:rmd17RMSEs}
\end{figure}

\begin{figure}[htp]
    \centering
    \includegraphics[width=0.75\linewidth]{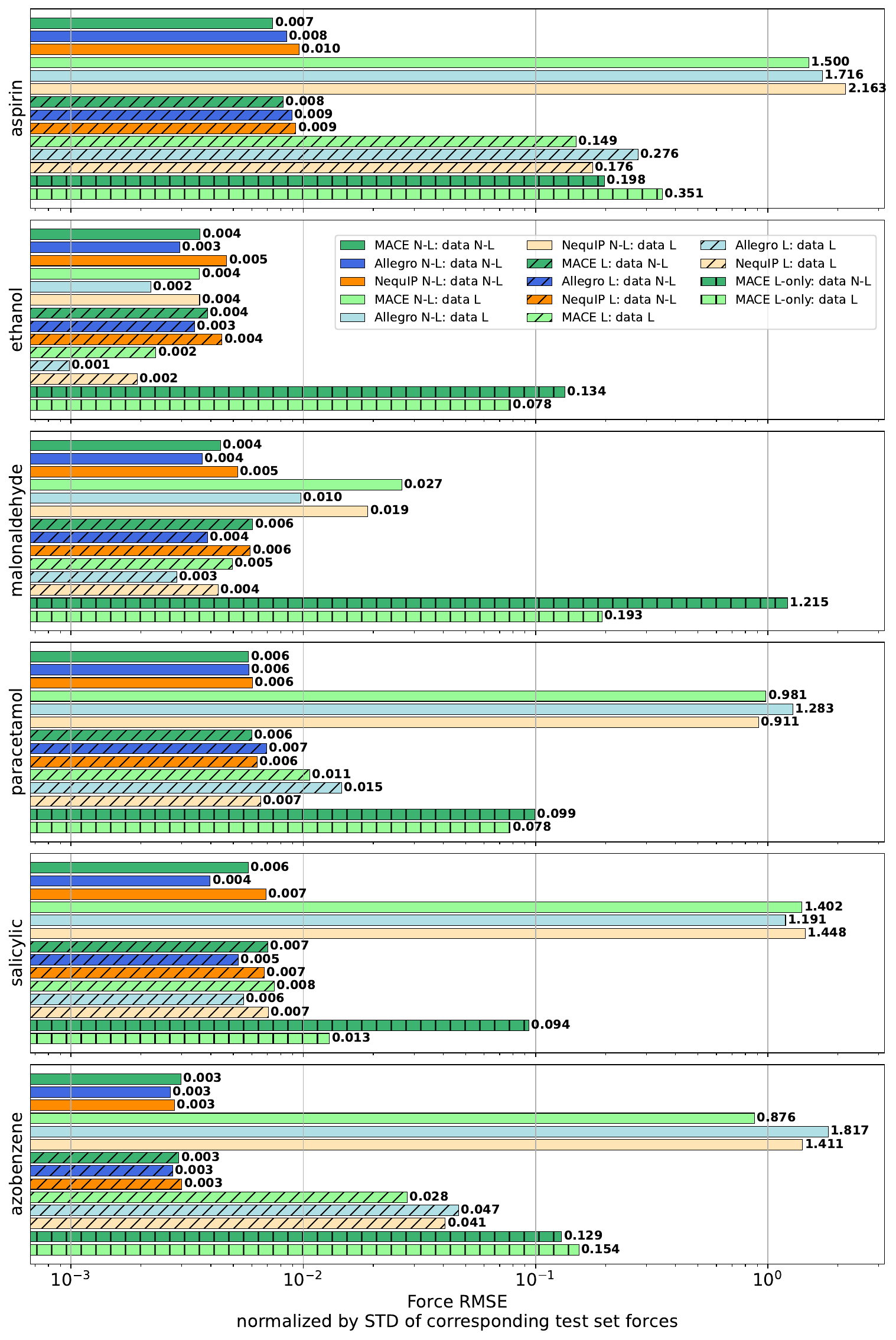}
    \caption{Force RMSEs normalized by the standard deviation of the target distribution.}
    \label{fig:rmd17ForcesRMSEs}
\end{figure}

Detailed bar charts visually comparing errors across models and datasets are
presented in Figs.~\ref{fig:rmd17RMSEs} and \ref{fig:rmd17ForcesRMSEs}. Note that
these charts compare the relative errors: the RMSE values divided by the
standard deviation of the target values.

Figs \ref{fig:phys-and-non-phys-allmolecs} and
\ref{fig:all_exact_matches_comparisons} show the numbers of physical and
unphysical stationary points, and matched and unmatched stationary points
respectively, for every molecule in the dataset. These figures extend Figs
\ref{fig:mlpLandscapes} \textbf{a, b} and \ref{fig:comparison} \textbf{a, b} in
the main text.

Unlike the original papers, we train our models with float64 precision, as some
authors later concluded that this approach is required to obtain a smooth
landscape.\cite{designSpaceMace} We also use gradient clipping of norm 10 as we
found it essential for low energy and force errors on the rMD17 test set.

\begin{table}[htbp]
\tiny
\centering
\caption{\label{table:MAE_and_RMSE_for_non_altitude_training} 
Energy (E) and force (F) errors based on mean absolute error (MAE) and root
mean square error (RMSE) for models trained on a Non-Landscape dataset,
reported in units of [meV] and [meV/Å], respectively, computed for hold-out
test set of 1,000 Non-Landscape configurations from rMD17. Values given in
parenthesis refer to the values reported in the original references, where the
test set contained 100,000 configurations.}
\begin{tabular}{lccccccccccccc}
\hline
 \multicolumn{2}{c}{\textbf{Model}} & \multicolumn{2}{c}{\textbf{Aspirin}} & \multicolumn{2}{c}{\textbf{Ethanol}} & \multicolumn{2}{c}{\textbf{Malonaldehyde}} & \multicolumn{2}{c}{\textbf{Paracetamol}} & \multicolumn{2}{c}{\textbf{Salicylic acid}} & \multicolumn{2}{c}{\textbf{Azobenzene}}\\ 
   &    & {MAE}  & {RMSE}  & {MAE}  & {RMSE} & {MAE}  & {RMSE}  & {MAE}  & {RMSE} & {MAE}  & {RMSE} & {MAE}  & {RMSE} \\ \hline
    \multirow{2}{*}{\textbf{NequIP}} & E & 2.2 (2.3)  & 3.6  & 0.4  (0.4)& 0.7 & 0.6 (0.8) & 1.0 & 1.1 (1.4) & 1.6 & 0.7 (0.7) &  1.6 & 0.5 (0.7) & 0.8\\
                            & F & 8.1 (8.2) &  13.0 & 2.7 (2.8) & 5.8 &  4.3 (5.1) & 7.3 & 5.1 (5.9)  & 8.1 & 4.1 (4.0) &  9.5 & 2.4 (2.9) & 3.8\\    
    \multirow{2}{*}{\textbf{Allegro}} & E & 1.8 (2.3)  & 2.7 & 0.3 (0.4)& 0.5 & 0.4
 (0.6)& 0.7 & 1.0 (1.5) & 1.5 & 0.5 (0.9) & 0.8 & 0.5 (1.2) & 0.8\\
                            & F & 6.7 (7.3) & 11.4 & 1.8 (2.1) & 3.7 & 2.9 (3.6)& 5.2 & 4.5 (4.9)  & 7.8 & 3.0 (2.9)  &  5.5 & 2.2 (2.6) & 3.7\\    
    \multirow{2}{*}{\textbf{MACE}} & E & 2.2 (2.2)  & 3.2 &  0.4 (0.4)& 0.6 &  0.6 (0.8)& 0.9 &  1.2 (1.4) & 1.7 &  0.7 (0.9) & 1.4 & 0.6 (1.2) & 0.9 \\
                            & F & 6.2 (6.6) & 10.0 &  2.4 (2.1) & 4.5 & 3.8 (4.1)& 6.2 &   4.8 (4.8) & 7.8 &  4.0 (3.1) & 8.0 & 2.5 (3.0) & 4.1 \\    
                            \hline

\end{tabular}
\end{table}

\begin{table}[htbp]
\tiny
\centering
\caption{\label{table:MAE_and_RMSE_for_non_altitude_training_sd_testing}Energy
(E) and force (F) errors based on mean absolute error (MAE) and root mean
square error (RMSE) for models trained on a Non-Landscape dataset, reported in
units of [meV] and [meV/Å], respectively, computed for a hold-out test set of
Landscape configurations (with number of configurations in the
paranthesis - cf. Table \ref{tab:molecule_data_sd_md_splits}).}
\begin{tabular}{lccccccccccccc}
\hline
 \multicolumn{2}{c}{\textbf{Model}} & \multicolumn{2}{c}{\textbf{Aspirin (1000)}} & \multicolumn{2}{c}{\textbf{Ethanol (101)}} & \multicolumn{2}{c}{\textbf{Malonaldehyde (290)}} & \multicolumn{2}{c}{\textbf{Paracetamol (1000)}} & \multicolumn{2}{c}{\textbf{Salicylic acid (729)}} & \multicolumn{2}{c}{\textbf{Azobenzene (365)}}\\ 
   &    & {MAE}  & {RMSE}  & {MAE}  & {RMSE} & {MAE}  & {RMSE}  & {MAE}  & {RMSE} & {MAE}  & {RMSE} & {MAE}  & {RMSE} \\ \hline
    \multirow{2}{*}{\textbf{NequIP}} & E & 517.3 & 579.6 & 0.0 & 0.1 & 0.1 & 0.2 & 252.0 & 314.2 & 85.6 & 119.0 & 327.4 & 452.0\\
                            & F & 146.9 & 302.5 & 0.4 & 0.6 & 1.1 & 1.5 & 52.3 & 127.4 & 176.7 & 364.5 & 60.2 & 181.7 \\    
    \multirow{2}{*}{\textbf{Allegro}} & E & 566.5 & 614.6 & 0.0 & 0.0  & 0.1 & 0.1 & 600.5 & 745.3 & 160.1 & 200.3 & 1290.2  & 1760.6 \\
                            & F & 131.1 & 240.1 & 0.3 & 0.4 & 0.5 & 0.7  & 80.8 & 179.3 & 162.2 & 299.7 & 95.9 & 234.1\\    
    \multirow{2}{*}{\textbf{MACE}} & E & 492.3 & 546.9 & 0.0 & 0.1 & 0.3 & 0.3 & 271.3 & 337.5 & 127.4 & 158.9 & 412.7 & 565.5 \\
                            & F & 107.7 & 209.8 & 0.4 & 0.6 & 1.4& 2.1 & 53.2 & 137.2 &178.3 & 352.8 & 47.3 & 112.9 \\    
                            \hline

\end{tabular}
\end{table}

\begin{table}[htbp]
\tiny
\centering
\caption{\label{table:MAE/_and_RMSE_for_altitude_training}
Energy (E) and force (F) errors based on mean absolute error (MAE) and root
mean square error (RMSE) for models trained on Non-Landscape and Landscape
datasets, reported in units of [meV] and [meV/Å], respectively, computed for a
hold-out test set of 1,000 Non-Landscape configurations from rMD17. Values
given in parenthesis refer to the values reported in the original references,
where the test set contained 100,000 configurations.}
\begin{tabular}{lccccccccccccc}
\hline
 \multicolumn{2}{c}{\textbf{Model}} & \multicolumn{2}{c}{\textbf{Aspirin}} & \multicolumn{2}{c}{\textbf{Ethanol}} & \multicolumn{2}{c}{\textbf{Malonaldehyde}} & \multicolumn{2}{c}{\textbf{Paracetamol}} & \multicolumn{2}{c}{\textbf{Salicylic acid}} & \multicolumn{2}{c}{\textbf{Azobenzene}}\\ 
   &    & {MAE}  & {RMSE}  & {MAE}  & {RMSE} & {MAE}  & {RMSE}  & {MAE}  & {RMSE} & {MAE}  & {RMSE} & {MAE}  & {RMSE} \\ \hline
    \multirow{2}{*}{\textbf{NequIP}} & E & 2.1 (2.3)  & 3.0  & 0.4 (0.4)& 0.7 &  0.7 (0.8) & 1.1 & 1.2 (1.4) & 1.7 & 0.8 (0.7) & 1.4  & 0.6 (0.7) & 0.8\\
                            & F &  7.7 (8.2) & 12.5  & 2.6 (2.8) & 5.6 &  4.8 (5.1) &  8.3 &  5.2 (5.9)  & 8.4 &  4.4 (4.0) &  9.4 & 2.5 (2.9) & 4.1\\    
    \multirow{2}{*}{\textbf{Allegro}} & E & 1.9 (2.3)  & 2.8 &  0.3 (0.4)& 0.5 & 0.4
 (0.6)& 0.8 & 1.0 (1.5) & 1.5 &  0.6 (0.9) & 1.1 & 0.5 (1.2) & 0.8\\
                            & F & 6.8 (7.3) & 12.1 & 1.8 (2.1) & 4.3 &  2.9 (3.6)& 5.5 &  4.7 (4.9)  & 9.3 & 3.2 (2.9)  & 7.2  & 2.2 (2.6) & 3.7\\    
    \multirow{2}{*}{\textbf{MACE}} & E & 2.5 (2.2)  & 3.6 &  0.4 (0.4)& 0.7 &  0.7 (0.8)& 1.1 & 1.7 (1.4) & 2.2 &   0.8 (0.9) & 1.1 &  0.7 (1.2) &  0.9 \\
                            & F & 6.9 (6.6) & 11.1 &  2.4 (2.1) & 4.8 & 5.0 (4.1)& 8.5 &  5.1 (4.8) & 8.0 &   4.4 (3.1) & 9.7 & 2.5 (3.0) & 4.0 \\    
                            \hline

\end{tabular}
\end{table}

\begin{table}[htbp]
\tiny
\centering
\caption{\label{table:MAE_and_RMSE_for_altitude_training_sd_testing}Energy (E)
and force (F) errors based on mean absolute error (MAE) and root mean square
error (RMSE) for models trained on Non-Landscape and Landscape datasets,
reported in units of [meV] and [meV/Å], respectively, computed for a hold-out
test set of Landscape configurations (with number of configurations
in the paranthesis - cf. Table \ref{tab:molecule_data_sd_md_splits}).}
\begin{tabular}{lccccccccccccc}
\hline
 \multicolumn{2}{c}{\textbf{Model}} & \multicolumn{2}{c}{\textbf{Aspirin (1000)}} & \multicolumn{2}{c}{\textbf{Ethanol (101)}} & \multicolumn{2}{c}{\textbf{Malonaldehyde (290)}} & \multicolumn{2}{c}{\textbf{Paracetamol (1000)}} & \multicolumn{2}{c}{\textbf{Salicylic acid (729)}} & \multicolumn{2}{c}{\textbf{Azobenzene (365)}}\\ 
   &    & {MAE}  & {RMSE}  & {MAE}  & {RMSE} & {MAE}  & {RMSE}  & {MAE}  & {RMSE} & {MAE}  & {RMSE} & {MAE}  & {RMSE} \\ \hline
    \multirow{2}{*}{\textbf{NequIP}} & E & 15.4 & 22.3 & 0.0 & 0.0 & 0.2 &0.2 & 0.4 & 0.5 &  0.4  & 0.5 & 0.4  & 0.6  \\
                            & F & 10.9 & 24.6 & 0.2 & 0.3 & 0.2& 0.3 & 0.4  & 0.9 & 0.8 & 1.8 & 0.4 & 0.6 \\    
    \multirow{2}{*}{\textbf{Allegro}} & E & 23.4 & 35.5 & 0.0 & 0.0 & 0.1 & 0.1 & 0.1 & 0.2 &  0.1  & 0.3 & 0.6 & 3.3 \\
                            & F & 15.2 & 38.6 & 0.1 & 0.2  & 0.2 & 0.2  &  0.3  & 2.0 & 0.6 & 1.4  & 0.7 & 6.0 \\    
    \multirow{2}{*}{\textbf{MACE}} & E & 14.6 & 19.6 & 0.0 & 0.0 &  0.1 & 0.1 & 1.1 & 1.3 & 2.1 & 2.6  & 1.9 & 2.5 \\
                            & F & 9.6 & 20.9 & 0.3 & 0.4 & 0.3  & 0.4  &  0.7  & 1.5 & 1.0  & 1.9  & 0.7 &3.6\\    
                            \hline

\end{tabular}
\end{table}

\newpage
\null 
\clearpage 

\section{MLIP landscapes}
\label{appendix:mlpLandscapes}

\begin{figure}[htp] 
        \centering
    \includegraphics[width=0.66\linewidth]{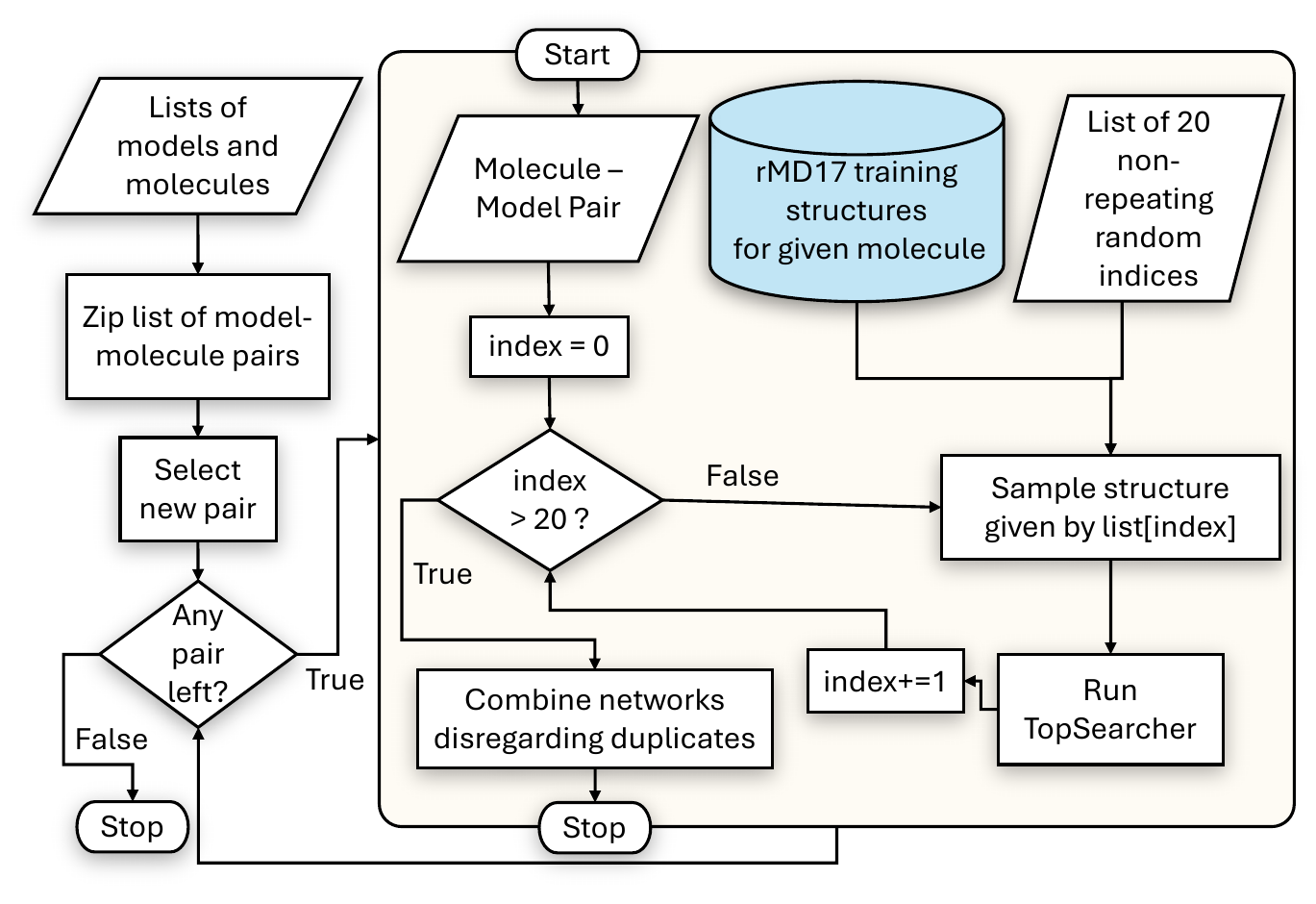}
        \caption{MLIP landscape generation flowchart.}
        \label{fig:flowchartLandscapeGen}
    \end{figure}

\begin{figure}[htp]
            \centering
        \includegraphics[width=0.5\linewidth]{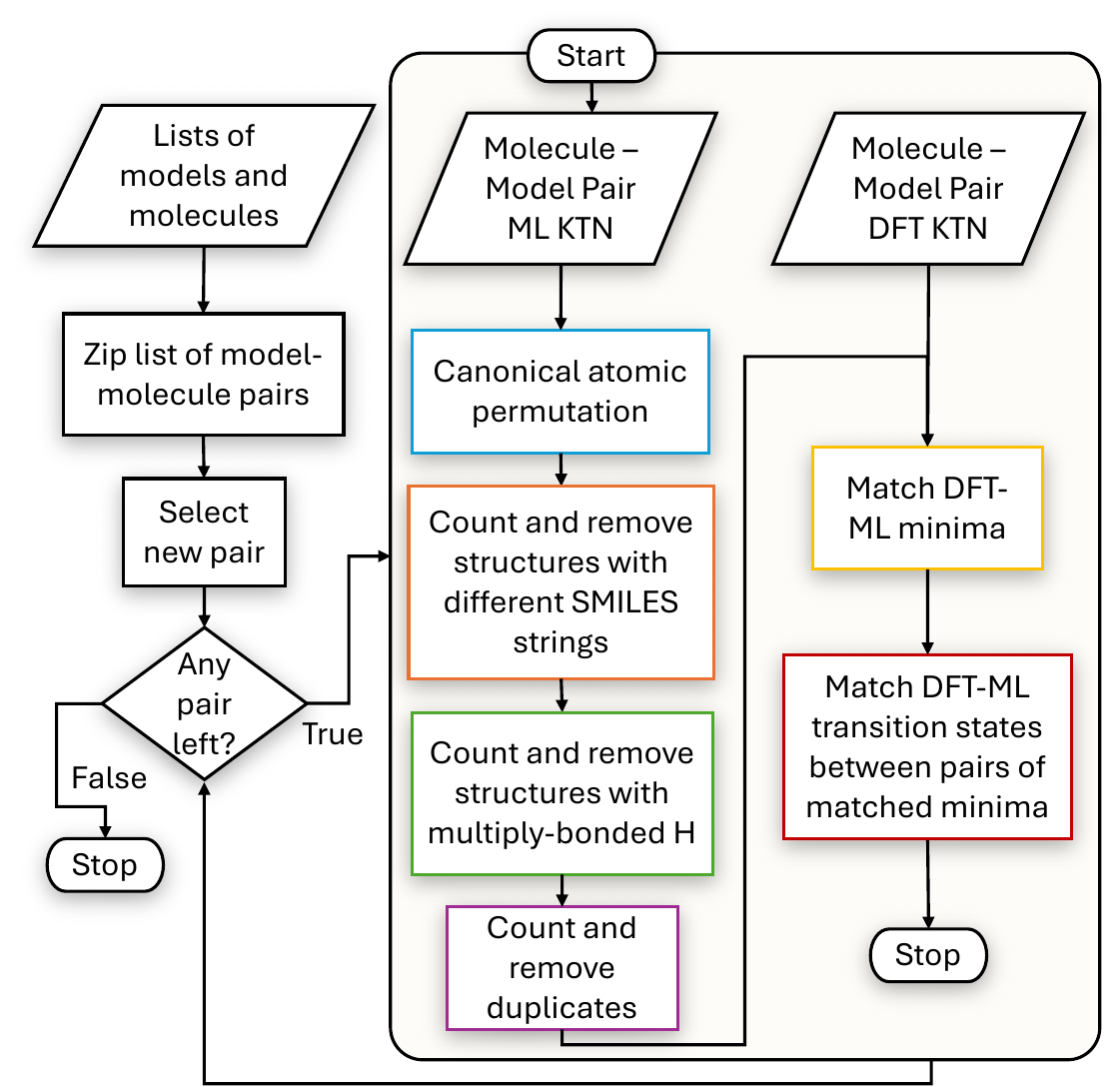}
	    \caption{Overview of the KTN comparison between DFT and MLIP (ML) runs.
The colors of the box outlines correspond to the detailed flowcharts shown
in Figs \ref{fig:flowchartSubmodules} and
\ref{fig:flowchartcomparisonAlgorithm}.}
            \label{fig:flowchartOverview}
        \end{figure}

\begin{figure}[htp]
        \centering
    \includegraphics[width=0.83\linewidth]{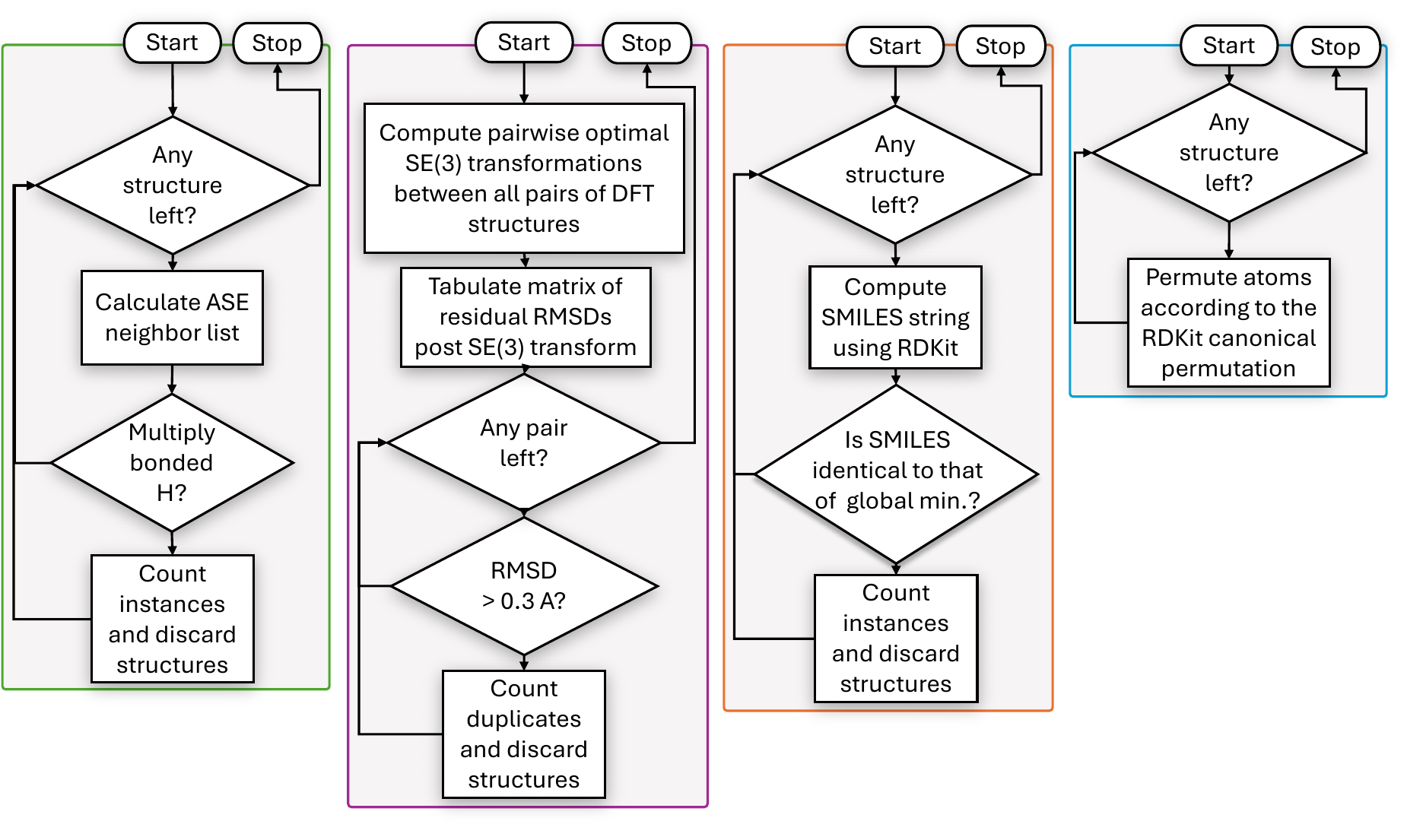}
        \caption{Submodules of Fig. \ref{fig:flowchartOverview}.}
        \label{fig:flowchartSubmodules}
    \end{figure}

\begin{figure}[htp]
        \centering
    \includegraphics[width=0.73\linewidth]{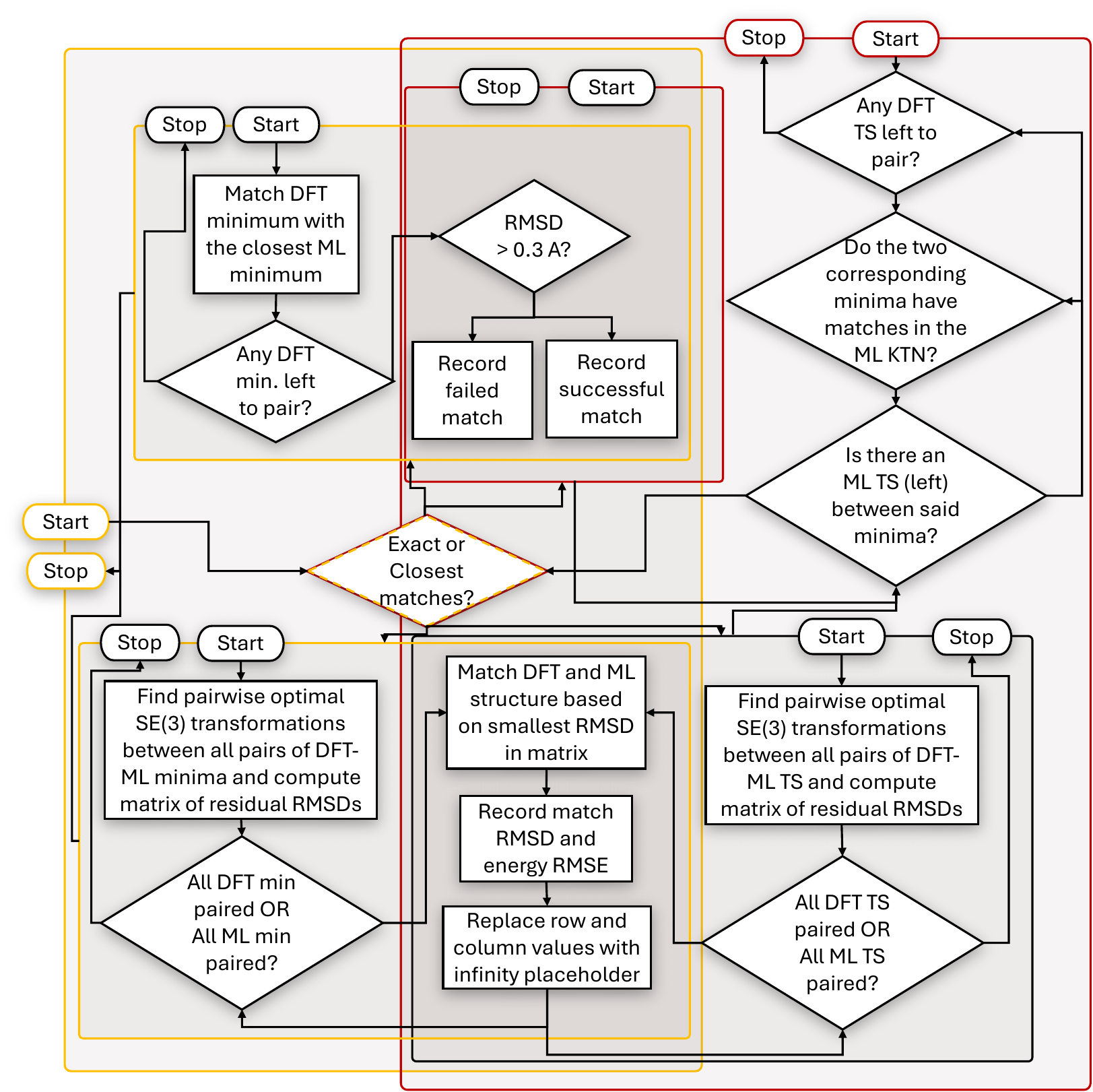}
        \caption{Details of the comparison algorithm with its two tracks: exact comparison and closest comparison. ML is shorthand for MLIP.}
        \label{fig:flowchartcomparisonAlgorithm}
    \end{figure}

\begin{figure}
    \centering
    \includegraphics[width=1\linewidth]{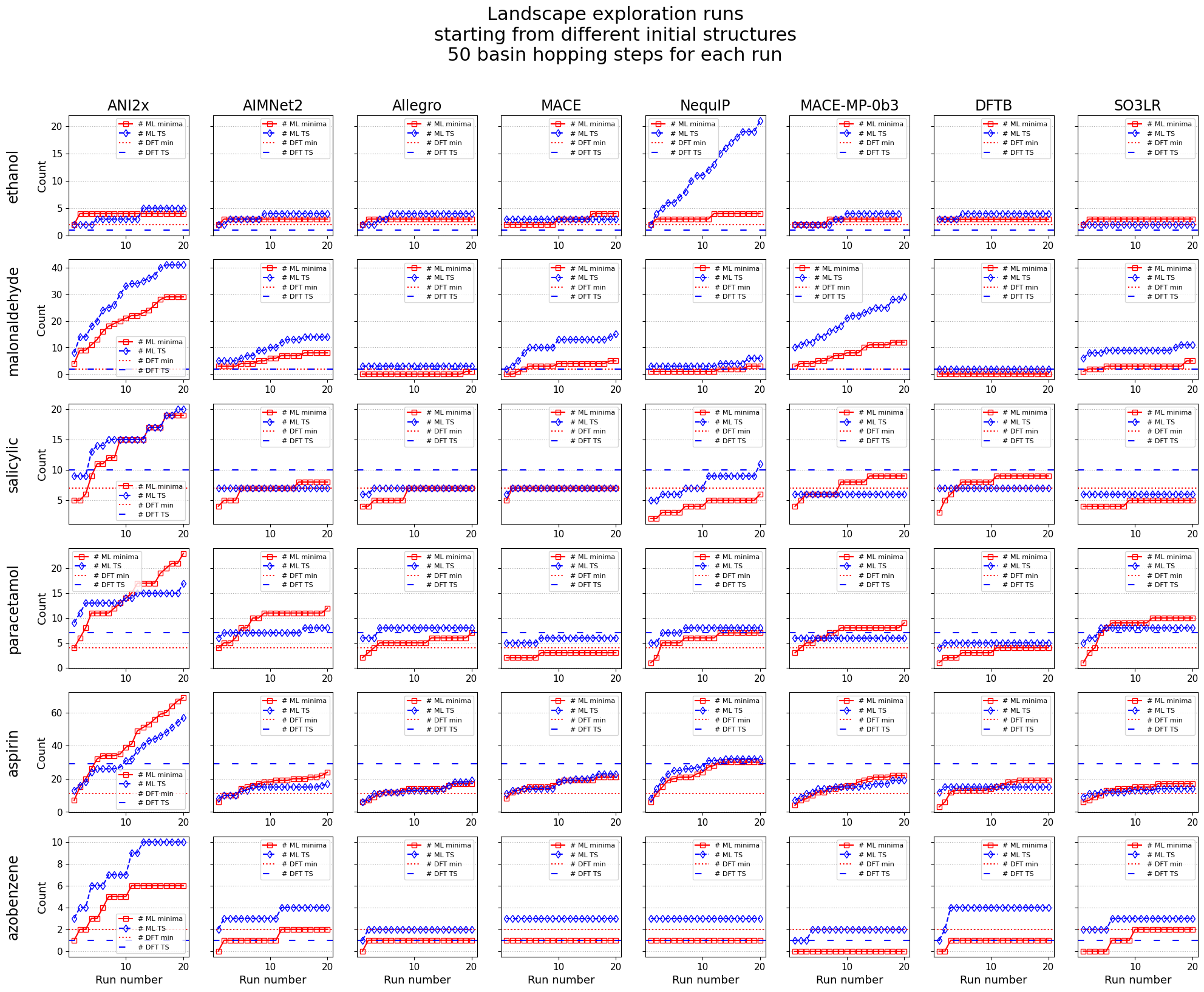}
    \caption{Increasing number of minima/TS with increasing number of landscape exploration runs for N-L models. ML is shorthand for MLIP.}
    \label{fig:evolutionof_n_ts_n_min}
\end{figure}

\begin{figure}
    \centering
    \includegraphics[width=0.7\linewidth]{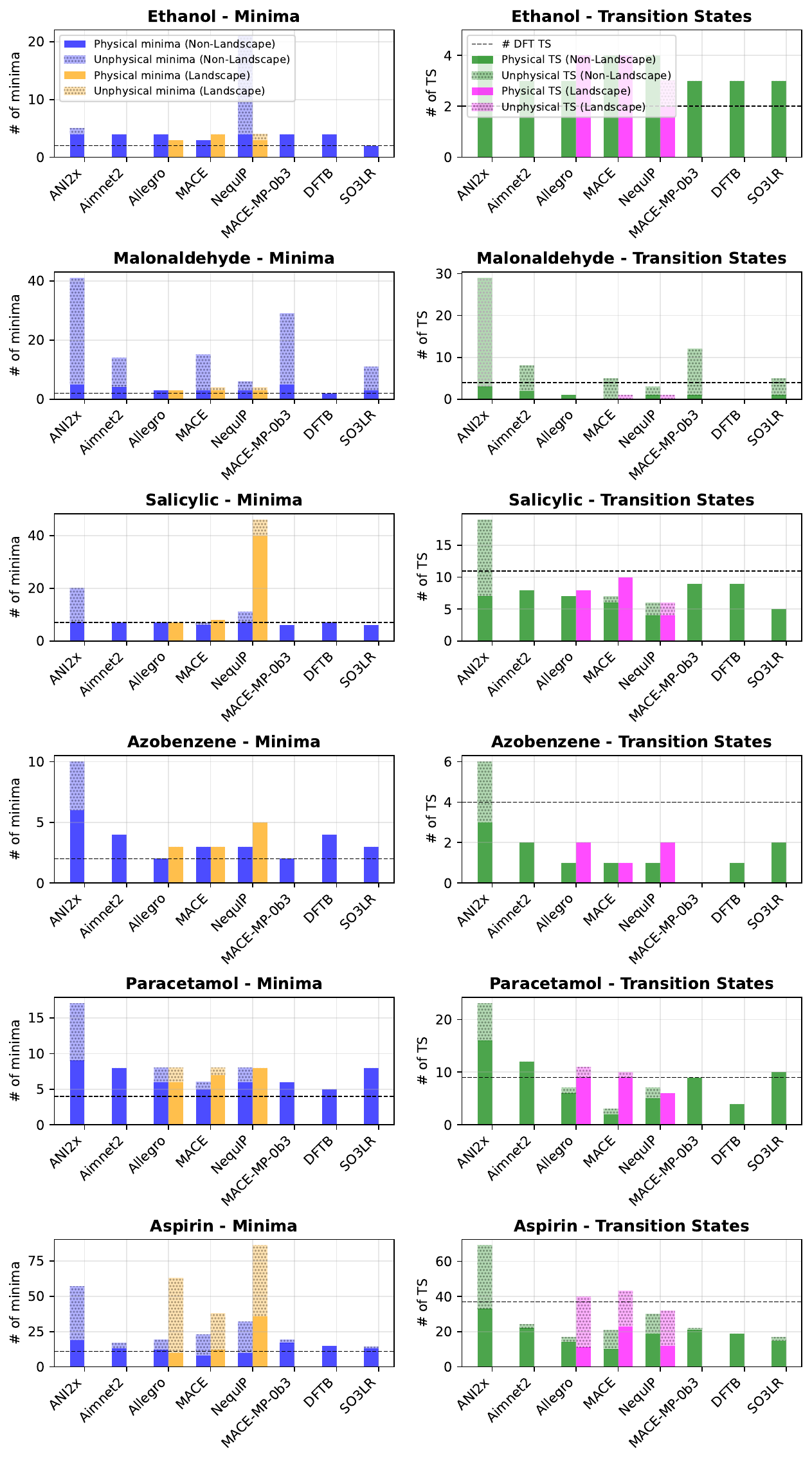}
    \caption{Numbers of physical and unphysical structures for each molecule, for every MLIP KTN. This figure extends Fig.~\ref{fig:mlpLandscapes} \textbf{a, b}.}
    \label{fig:phys-and-non-phys-allmolecs}
\end{figure}

\begin{figure}
    \centering
    \includegraphics[width=0.7\linewidth]{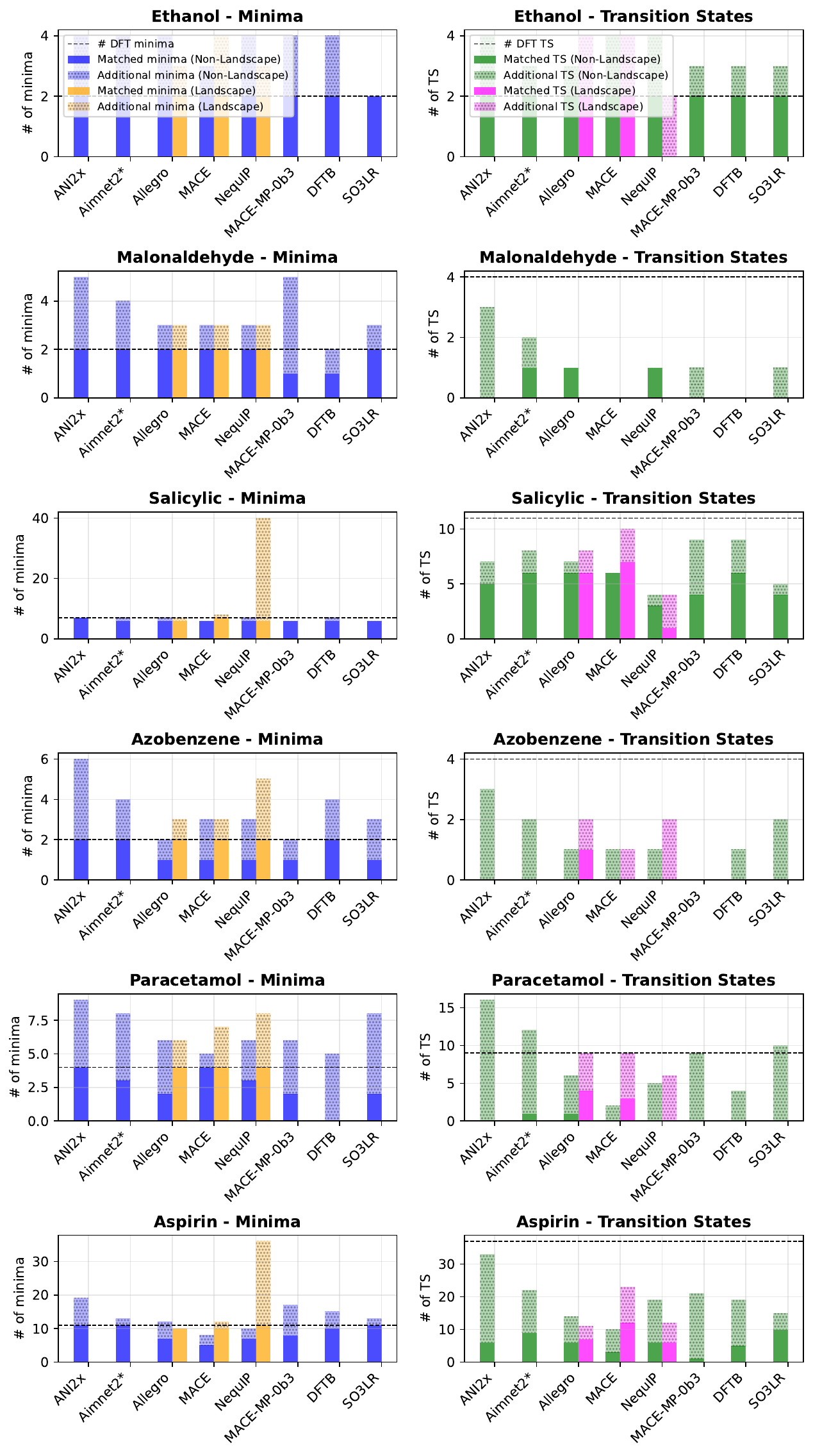}
    \caption{Numbers of matched and additional stationary points for each molecule, for each model. This figure extends Fig.~\ref{fig:comparison} \textbf{a, b}.}
    \label{fig:all_exact_matches_comparisons}
\end{figure}

Fig.~\ref{fig:flowchartLandscapeGen} show the workflow used to obtain the
MLIP landscapes and Figs.~\ref{fig:flowchartOverview} and
\ref{fig:flowchartSubmodules}, \ref{fig:flowchartcomparisonAlgorithm} provide
flowcharts illustrating the comparison of the DFT and MLIP KTNs.
Fig.~\ref{fig:evolutionof_n_ts_n_min} displays the evolution of the total
number of minima and transition states with increasing number of landscape
exploration runs for each molecule and for each model. The plot extends the
results shown in Fig. \ref{fig:comparison} \textbf{e}.

\end{document}